\documentclass[%
 reprint,
 amsmath,amssymb,
 aps,
 prc,
]{revtex4-2}

\usepackage{graphicx}
\usepackage{dcolumn}
\usepackage{bm}

\begin{document}

\title{Implementation of the Moments Method in TALYS: Sensitivity of $^{56,57}$Fe$(n,\gamma)$ observables to shell model nuclear level densities}%

\author{Lucas Chouinard}%
\affiliation{%
 Physics Department, Grand Valley State University, Allendale, MI, US
}%

\author{Sofia Karampagia}
 \email{karampso@gvsu.edu}
\affiliation{
 Physics Department, Grand Valley State University, Allendale, MI, US
}%
\affiliation{
 Facility for Rare Isotope Beams, Michigan State University, East Lansing, MI, US
}%

\begin{abstract}
The Moments Method (MM), a statistical spectroscopy framework built upon configuration interaction 
shell model Hamiltonians that enables shell model descriptions of nuclear level densities without 
requiring full diagonalization of the many body Hamiltonian, is implemented in the TALYS reaction 
code to investigate the sensitivity of Hauser-Feshbach calculations for the $^{56}$Fe$(n,\gamma)^{57}$Fe 
and $^{57}$Fe$(n,\gamma)^{58}$Fe reactions to the nuclear level density input. Spin- and parity-dependent 
MM level densities calculated in the pf shell are interfaced with TALYS and compared with the 
phenomenological and microscopic level density models available in the code. To construct TALYS ready inputs, 
the MM level densities are extended to higher excitation energies using a back shifted Fermi gas continuation 
together with an experimentally constrained prescription for opposite parity states. A sensitivity band for 
the MM calculations is estimated by varying a single parameter governing the statistical reconstruction of 
the level density. Neutron capture cross sections, Maxwellian averaged cross sections, and reaction rates are 
compared with evaluated ENDF/B-VIII.0 data and recommended KADoNiS and JINA-CEE values to assess the relative 
importance of the nuclear level density and $\gamma$-ray strength function inputs. The calculations show that 
the observables are considerably more sensitive to the choice of $\gamma$-ray strength function than to the 
MM level density variation considered here, while the MM-based results remain in good agreement with evaluated 
cross sections and recommended astrophysical reaction rates.
\end{abstract}

\maketitle


\section{\label{sec:intro}Introduction}
Neutron capture reactions play a central role in a wide range of topics in nuclear physics and astrophysics. 
In particular, the synthesis of elements heavier than iron in stellar environments proceeds primarily through 
neutron capture processes, such as the slow (s) \cite{Pignatari_2010, RevModPhys.83.157}, 
rapid (r) \cite{ARNOULD200797}, and intermediate (i) \cite{Cowan_1977} processes, where the competition between 
neutron capture and $\beta$ decay determines the evolution of isotopic abundances \cite{Sneden_2008}. Accurate 
neutron capture cross sections and reaction rates are therefore essential inputs for astrophysical 
reaction network calculations. Neutron captures are also important in nuclear technology applications, 
including reactor design \cite{Gelles_1990, TERRANI2014420, Rebak_2017}, nuclear waste transmutation \cite{Colonna_2010, NEA_2002}, 
and isotope production. \cite{Knapova_25}. 

So far, the experimental pathway toward determining neutron capture rates for neutron rich isotopes is through 
indirect measurements, such as the Oslo method \cite{Larsen_2019, Guttormsen_1987, Schiller_2000}, the inverse 
Oslo method \cite{Ingeberg_2000, Ingeberg_2025} and the $\beta$-Oslo method \cite{Spyrou_2014, Liddick_2016, Larsen_2019} 
and surrogate techniques \cite{Escher_2012, Escher_2018, Ratkiewicz_2019}. In place of experimental measurements, 
neutron capture reaction cross sections and rates are estimated within the statistical Hauser-Feshbach model \cite{Hauser}, 
where the calculated cross sections depend sensitively on nuclear statistical properties such as the nuclear level density 
(NLD) and the $\gamma$-ray strength function ($\gamma$SF). These two ingredients introduce the largest uncertainties in 
neutron capture calculations. The $\gamma$SF describes the average probability of emitting or absorbing a $\gamma$-ray 
of a given energy in a nucleus. The NLD is the number of available energy levels per unit energy interval at a given 
excitation energy. 

Since Bethe's pioneering work in 1936 \cite{Bethe_1936}, which attempted to calculate the number of energy levels of a heavy 
nucleus near neutron separation excitation energy, several functional forms for the estimation of NLDs have been 
proposed (see references in \cite{Capote_2009}). Among these, traditional reaction codes, such as TALYS, use phenomenological 
models such as the composite Gilbert Cameron model \cite{Gilbert_1965}, the back shifted Fermi gas model \cite{Dilg_1973} and 
the generalized Superfluid model \cite{Ignatyuk_1993}. These rely on parameters fitted to experimental data or deduced from 
systematics, limiting their reliability to nuclei close to the valley of stability. Microscopic approaches implementing 
Skyrme and Gogny energy-density functionals within the Hartree-Fock-Bogoliubov (HFB) framework \cite{Goriely_2001, Goriely_2008, Hilaire_2012} 
provide spin- and parity- dependent global NLD predictions across the nuclear chart and are implemented in TALYS \cite{TALYS}. 

Configuration interaction shell model calculations using conventional diagonalization could also be implemented to derive 
spin- and parity- dependent NLDs by counting the number of energy levels per energy interval, however, the calculations 
become intractable already a few nucleons away from the $^{40}$Ca core. Shell model Monte Carlo methods have proven to be a successful 
alternative to microscopic calculations of spin- and parity- dependent NLDs \cite{Alhassid_2015} and $\gamma$SF \cite{DeMartini_2025} 
for medium mass and heavy nuclei. Although these calculations demonstrate agreement with available experimental data \cite{Alhassid_2015} 
they cannot be implemented for global studies to fully replace the HFB microscopic approaches. Shell model approaches that 
compute NLDs from the moments of the Hamiltonian have demonstrated that they are in good agreement with conventional 
configuration interaction shell model calculations \cite{Ormand_2020, Senkov_2016}. 

In this paper we perform a sensitivity study of Moments Method (MM) \cite{Senkov_2016} nuclear
level densities in the TALYS reaction code and investigate their impact on the calculated cross
sections and Maxwellian averaged reaction rates for the $^{56}$Fe$(n,\gamma)^{57}$Fe and
$^{57}$Fe$(n,\gamma)^{58}$Fe reactions. In Section~\ref{sec:MM} the MM is described. In
Section~\ref{sec:LD} the calculated NLDs are compared with experimental data and the method used
to interface them with TALYS is described. In Section~\ref{sec:CS_MACS} the calculated cross
sections and Maxwellian averaged reaction rates are presented. The results are discussed in
Section~\ref{sec:D}, followed by conclusions in Section~\ref{sec:C}.


\section{\label{sec:MM}Moments Method}
The approach used in this work to calculate nuclear level densities is based on methods of 
statistical spectroscopy \cite{French1, French2, French3, Wong}, employing effective shell model 
Hamiltonians. Further details on the method and its applications can be found in Refs. 
\cite{HoroiM1, HoroiM2, HoroiM3, Senkov, Scott, Senkov1, Senkov_2016}. Here, we summarize its main 
features. 

The moments method is based on configuration interaction shell model ideas and provides an efficient 
alternative to full diagonalization. Instead of diagonalizing large Hamiltonian matrices, it reconstructs 
the many body level density from the first two moments, namely, the centroid and width, of the Hamiltonian 
in each partition. The shell model space is decomposed into all possible partitions $p$, corresponding to 
the different ways in which valence particles can occupy the available single particle orbitals. It is 
assumed that the level density associated with each partition can be approximated by a Gaussian distribution. 
This assumption is supported by studies of statistical spectroscopy \cite{Brody}, as well as by results 
obtained from exact shell model diagonalizations.

The total level density as a function of excitation energy $E$, $\rho(E)$, is then expressed as a sum 
over partitions:

\begin{equation}
  \rho(E;a)=\sum_p D_{ap}G_{ap}(E), \label{1}
\end{equation} 
where $G_{ap}$ is a finite range Gaussian describing partition $p$ for a given set of quantum numbers $a$, 
such as spin $J$, isospin projection $T_z$ and parity $\pi$, and $D_{ap}$ is the corresponding dimension 
(number of states). 

The finite range Gaussians are defined as
\begin{equation}
  G_{ap}(E) = G(E-E_{ap}+E_{g.s.};\sigma_{ap}), \label{2}
\end{equation}
with
\begin{equation}
G(x;\sigma) = C
  \begin{cases}
    e^{-x^2/2\sigma^2},       & |x| \leq \eta \sigma,\\
    0,                        & |x| > \eta \sigma                            \label{3}
  \end{cases}
\end{equation}
and $C$ is a normalization constant satisfying $\int dx \ G(x;\sigma)=1$, and $\eta$ is a cutoff 
parameter, typically in the 2.5-3.0 \cite{Senkov, Scott}. 

The characteristics of each Gaussian, namely its centroid $E_{ap}$ and width $\sigma_{ap}$, 
are determined by the moments (traces) of the effective shell model Hamiltonian. The centroid 
corresponds to the first moment, 
\begin{equation}
E_{ap} = \langle H \rangle_{ap} = \frac{1}{D_{ap}}Tr^{ap}H,  \label{4}
\end{equation}
while the width is given by the second moment, 
\begin{equation}
\sigma_{ap}^2 = \langle H^2 \rangle_{ap} - E_{ap}^2 = \frac{1}{D_{ap}}Tr^{ap}H^2 - E_{ap}^2.  \label{5}
\end{equation}  
An important ingredient for the correct positioning of the Gaussian distribution on the 
excitation energy axis is the ground state energy $E_{g.s.}$. This can be obtained  
either from shell model diagonalization, when feasible, or from the exponential 
convergence method \cite{Horoi1, Horoi2, Horoi3}. In addition to the ground state energy, 
knowledge of the yrast energies is beneficial for constraining the low energy behavior of the 
level density. The yrast energy can be obtained from shell model calculations or experimental data 
when available. 

The moments method has been further improved by incorporating an algorithm that removes spurious 
center of mass excitations when multi shell model spaces are considered \cite{Horoi4}. At present, 
the method can be readily applied to nuclei in model spaces ranging from the $sd$ up to 
the $pf+g_{9/2}$ model space.


\section{\label{sec:LD}Level Densities of $^{56,57}$Fe}

For the calculation of the $^{56,57}$Fe level densities we employ the $pf$ model space with a $^{40}$Ca 
core and the GXPF1A Hamiltonian \cite{Honma2004, Honma2005}. The ground state energies are calculated 
using the NuShellX code \cite{NuShellX}, while yrast energies for the lowest spin states  
are taken from experimental data \cite{NNDCFe}. For the cutoff parameter $\eta$, we consider values in 
the range 2.3 $\le \eta \le$ 2.6. This range is motivated by a study of the possible dependence of 
$\eta$ on spin and excitation energy. Specifically, we performed exact shell model calculations using 
NuShellX for the first 200 states with spins $J^{\pi}=0^+-8^+$ in $^{64}$Fe and compared the resulting 
level densities with those obtained using the MM. We observed that $\eta$ decreases slowly with increasing 
spin and increases slowly with excitation energy. Due to computational limitations, a similar analysis 
cannot be carried out for $^{56,57}$Fe. Instead, we adopt a range of $\eta$ values that reflects the 
expected variation of MM level densities with spin and excitation energy. In the present work this range is interpreted 
as a model sensitivity band for the MM calculations rather than as a statistical confidence interval. This NLD sensitivity band is propagated through 
the Hauser-Feshbach calculations, enabling an assessment of its impact on neutron capture cross sections and 
Maxwellian averaged reaction rates.

\subsection{Comparison of Moments Method level densities to experimental data}

\begin{figure}
\centering
\includegraphics[height=60mm]{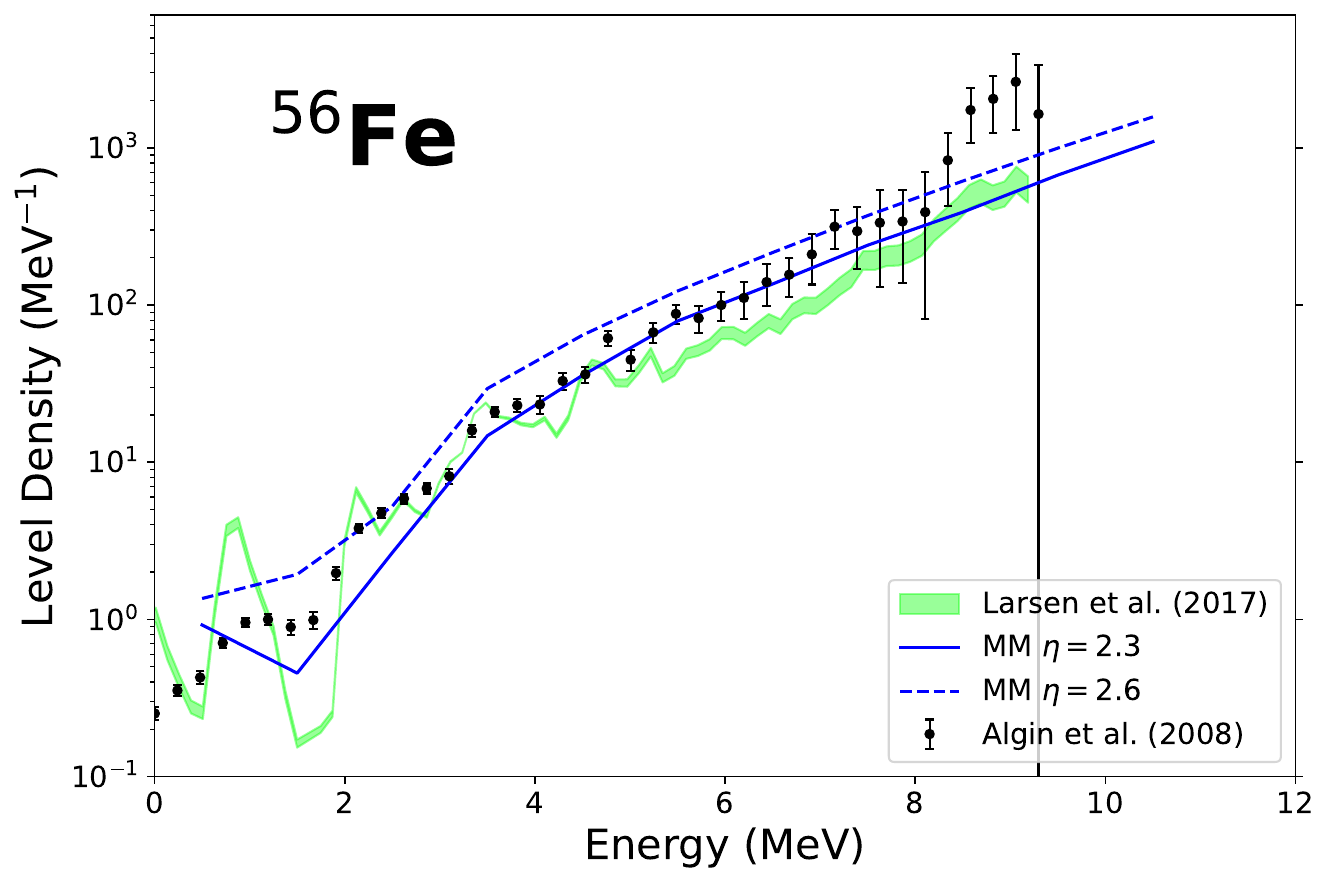}
\caption[]{Comparison of the cumulative level density of $^{56}$Fe calculated using the MM for $\eta$ = 2.3 (blue solid line) and $\eta$ = 2.6 (blue dashed line), for states $J^{\pi}=0^+-12^+$. The results are compared with level densities extracted using the Oslo method from two independent datasets: ($^3$He,$\alpha\gamma$) reactions (black points \cite{Algin}) and (p,p') reactions (green shaded region representing upper and lower normalization limits from Ref.~\cite{Larsen} and C.~Larsen, private communication).}\label{fig1}
\end{figure}

\begin{figure}
\centering
\includegraphics[height=60mm]{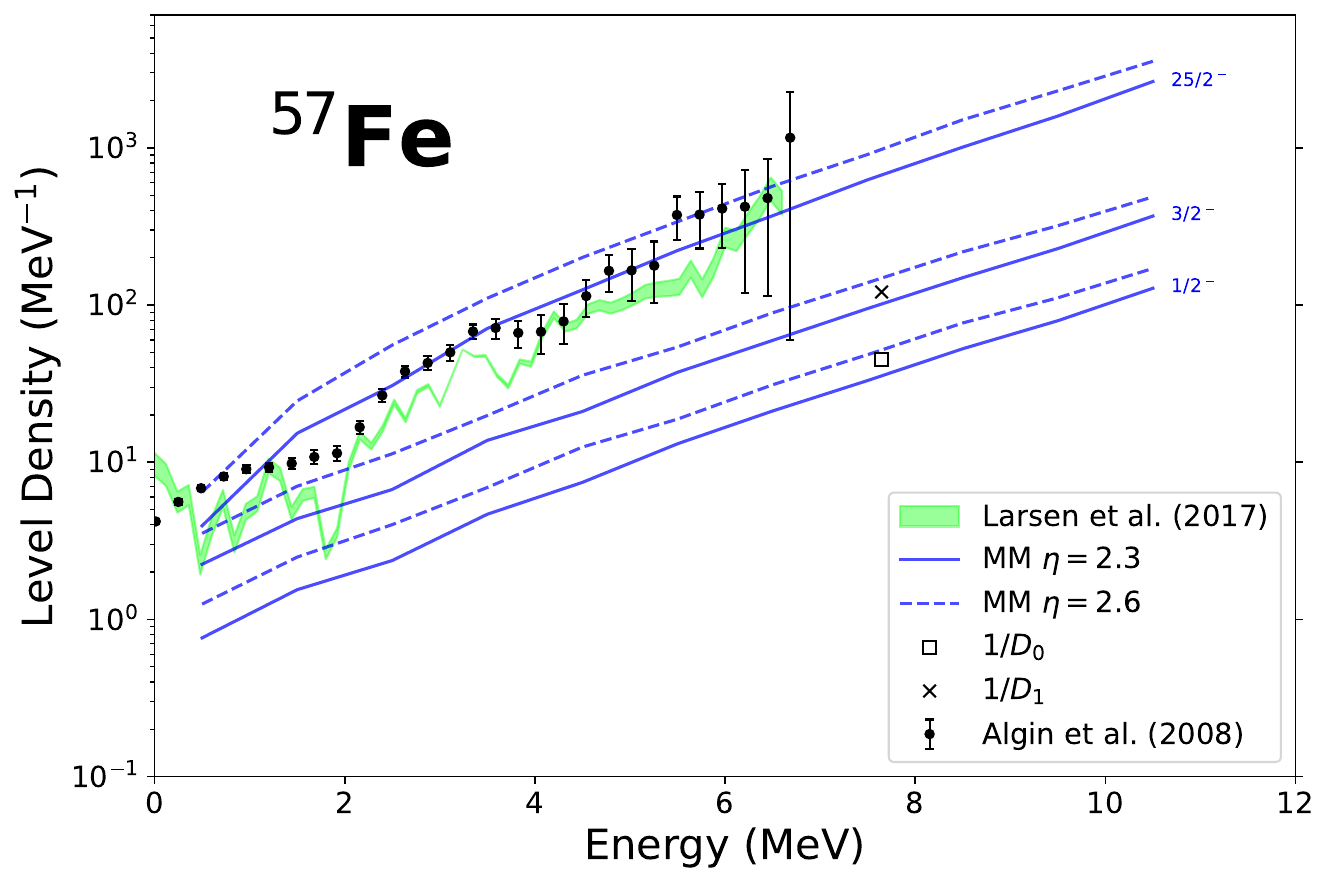}
\caption[]{Comparison of the cumulative level density of $^{57}$Fe calculated using the MM for $\eta$ = 2.3 (blue solid line) and $\eta$ = 2.6 (blue dashed line), for states $J^{\pi}=1/2^+-25/2^+$. The results are compared with level densities extracted using the Oslo method from two independent datasets: ($^3$He,$\alpha\gamma$) reactions (black points \cite{Algin}) and (p,p') reactions (green shaded region representing upper and lower normalization limits from Ref.~\cite{Larsen} and C.~Larsen, private communication). Also shown are the MM level densities at the neutron separation energy: the $J^{\pi}=1/2^-$ component compared with the s-wave value (square symbol) and the cumulative $J^{\pi}=1/2^--J^{\pi}=3/2^-$ component compared with the p-wave value (cross symbol).}\label{fig2}
\end{figure}

In Figs. \ref{fig1} and \ref{fig2} we compare the MM level densities with those extracted 
using the Oslo method. Because the experimental Oslo data are limited to a restricted spin 
range, we compute level densities for $J^{\pi}=0^+-12^+$ in $^{56}$Fe and $J^{\pi}=1/2^--25/2^-$ 
in $^{57}$Fe. The green shaded region represents Oslo data obtained from (p,p') reactions 
\cite{Larsen}, where the band reflects uncertainties associated with the normalization procedure.
The error bars include both statistical uncertainties and systematic contributions arising from 
the unfolding of $\gamma$-ray spectra and the extraction of primary $\gamma$ rays. 
These data typically correspond to a more restricted spin range compared to results from 
($^3$He, $\alpha\gamma$) reactions (black points) \cite{Algin}, which do not include systematic 
normalization uncertainties. This difference in spin population accounts for the observed 
discrepancies between the two experimental datasets across excitation energy. The MM level densities 
are shown as a blue solid line ($\eta=2.3$) and blue dashed line ($\eta=2.6$). Despite the statistical 
nature of the MM approach, the calculated results reproduce the overall trend of the Oslo data. 
For $^{57}$Fe, we also include the s-wave (square symbol) and p-wave (cross symbol) level densities at 
the neutron separation energy \cite{Mughabghab}. These values lie within the range spanned by the MM 
calculations, providing additional consistency with experimental constraints. 

\subsection{Implementation of Moments Method level densities in TALYS}

In order to use the Moments Method (MM) nuclear level densities in reaction calculations, the MM results 
were interfaced with the TALYS reaction code. This was achieved by constructing input level density tables 
formatted to mimic the parity dependent microscopic level density tables already implemented in TALYS, 
specifically reproducing the format of level density model 5 in TALYS, which is based on the 
Hartree-Fock-Bogoliubov plus combinatorial approach \cite{Goriely_2008}.

The MM provides spin- and parity-dependent level densities up to a certain excitation energy determined by 
the model space. In the present case, the $pf$ shell model space allows for a reliable description of states 
of a given parity up to moderate excitation energies. At higher excitation energies, contributions from 
configurations outside the model space become important, and an extrapolation of the MM level densities is 
required. To extend the MM level densities to higher excitation energies, we employ the back shifted Fermi gas 
(BSFG) model: 

\begin{equation}
  \rho(E_x,J)=\frac{\sqrt{\pi}}{12} \frac{\exp\left[2\sqrt{\alpha_J \left( E_x - \Delta_J \right)} \right]}{ \alpha_J^{1/4} \left( E_x - \Delta_J \right)^{5/4} } \label{6}
\end{equation} 
where $\alpha_J$ and $\Delta_J$ are treated as $J$-dependent empirical parameters obtained by fitting
the MM level densities in the matching region. 

The extrapolation is performed by matching the $J$-dependent MM level densities to the BSFG functional form at 
an energy where the MM calculation is still reliable, ensuring a smooth transition between the microscopic and 
phenomenological descriptions. This 
approach avoids the purely exponential behavior of the constant temperature model 
($\rho(E_x,J)=\frac{1}{T} \exp\left( \frac{E_x-E_0}{T}\right)$, with the nuclear temperature $T$ and $E_0$ parameters
to adjust the formula to the experimental discrete levels) at higher excitation energies 
and provides a more realistic asymptotic behavior of the level density.

The MM calculations in the $pf$ model space include only a single parity. Therefore, an additional prescription 
is required to account for opposite parity (intruder) states. In the absence of complete experimental constraints 
on the parity dependence of the level density, we construct the intruder parity contribution using available 
experimental level schemes from NNDC.

Specifically, we identify an excitation energy $E_0$ at which the first opposite parity (intruder) states appear. 
At this energy, the ratio of intruder to normal parity level densities is taken to be zero, 
$\rho_{\mathrm{intruder}}/\rho_{\mathrm{normal}} = 0$. We then determine an energy $E_1$ at which the populations 
of positive and negative parity states become comparable, corresponding to parity equipartition, 
$\rho_{\mathrm{intruder}}/\rho_{\mathrm{normal}} = 1$. Above $E_1$, parity equipartition is assumed, such that 
positive  and negative parity states contribute equally to the 
total level density.

Between $E_0$ and $E_1$, the ratio $\rho_{\mathrm{intruder}}/\rho_{\mathrm{normal}}$ is assumed to evolve smoothly 
with excitation energy, following a sigmoid-like behavior consistent with the gradual onset of intruder configurations \cite{Sangeeta_2022}:
\begin{equation}
\frac{\rho_{\mathrm{intruder}}}{\rho_{\mathrm{normal}}}
= \frac{1}{1 + \exp\left[-\beta \left(E_x - E_c\right)\right]},
\label{eq:parity}
\end{equation}
where $\beta$ and $E_c$ are parameters determined by requiring consistency with the boundary conditions at $E_0$ 
and $E_1$.

We note that the total level density is only weakly sensitive to the precise choice of the equilibration energy $E_1$. 
In the present sensitivity study, however, the sensitivity band shown below reflects only the variation of the MM cutoff parameter $\eta$ and does not include an additional variation of the opposite parity prescription. The resulting spin- and parity-dependent level densities, including the BSFG extrapolation and 
the treatment of opposite parity states, are then provided to TALYS (level density model 5 \cite{Goriely_2008}) and 
used as input for the Hauser-Feshbach calculations of neutron capture cross sections and Maxwellian averaged reaction 
rates.


\section{\label{sec:CS_MACS}($n,\gamma$) Cross sections and Maxwellian averaged reaction rates}

In this section, we investigate the impact of the MM level densities on the $(n,\gamma)$ cross sections 
and Maxwellian averaged reaction rates and compare the results with a range of models available in TALYS.

The $(n,\gamma)$ cross sections and Maxwellian averaged reaction rates for the $^{56}$Fe$(n,\gamma)$ and 
$^{57}$Fe$(n,\gamma)$ reactions were calculated using version 2.0 of the TALYS reaction code by systematically exploring 
all available nuclear level density (NLD) and $\gamma$-ray strength function ($\gamma$SF) models. These calculations 
were then juxtaposed with results obtained using the MM level densities combined with all available $\gamma$SF 
parameterizations, allowing for a direct assessment of the sensitivity of the observables to the underlying 
nuclear statistical properties.

In TALYS, the NLD models considered include both phenomenological and microscopic approaches. The 
phenomenological models correspond to (i) the Constant Temperature and Fermi gas (CTM) model \cite{Gilbert_1965}, 
(ii) the back shifted Fermi gas (BSFG) model \cite{Dilg_1973}, and (iii) the Generalized Superfluid Model (GSM) 
\cite{Ignatyuk_1993}, while the microscopic models include level densities based on Hartree-Fock-Bogoliubov (HFB) 
plus combinatorial calculations \cite{Goriely_2001, Goriely_2008, Hilaire_2012}. The $\gamma$-ray strength functions 
considered include different Lorentzian parameterizations of the E1 strength, such as the standard Lorentzian (SLO) 
\cite{Brink_1957, Axel_1962}, generalized Lorentzian (GLO) \cite{Kopecky_1941}, and the hybrid model of Goriely 
with energy- and temperature-dependent width \cite{Goriely_1998} and various microscopic $\gamma$SF models 
\cite{Goriely_2002, Goriely_2004, Daoutidis_2012, Goriely_2018}. For the optical model potential, we employ 
the TALYS default, namely the Koning and Delaroche optical model potential \cite{Koning_2003}, as it has been 
previously shown that, for this type of calculation, variations in the NLD and $\gamma$SF dominate the sensitivity of $(n,\gamma)$ observables \cite{Liddick_2016}.

\begin{figure}
\centering
\includegraphics[height=60mm]{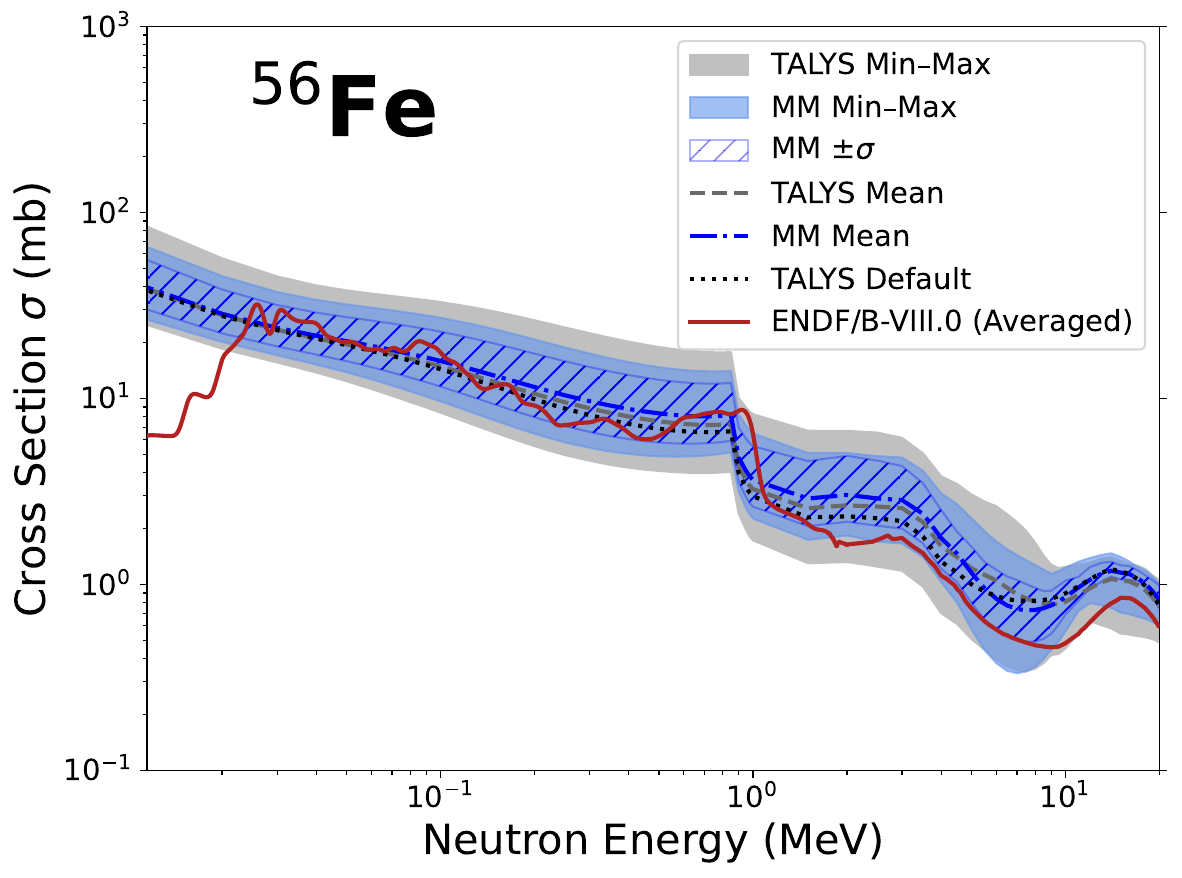}
\caption[]{Neutron capture cross section for $^{56}$Fe$(n,\gamma)$ as a function of neutron energy, 
calculated with TALYS. The gray band shows the range (min-max) obtained from all combinations 
of NLD and $\gamma$SF models available in TALYS, while the blue band corresponds 
to the full spread obtained using the MM level densities combined with all $\gamma$SF models. 
The hatched region represents the central 68\% interval of the MM-based results. The dashed and 
dash-dotted lines denote the TALYS and MM mean values, respectively, while the dotted line shows 
the TALYS default calculation. Evaluated data from the ENDF/B-VIII.0 library \cite{Brown2018} are 
shown for comparison (red solid line).}\label{fig3}
\end{figure}

\begin{figure}
\centering
\includegraphics[height=60mm]{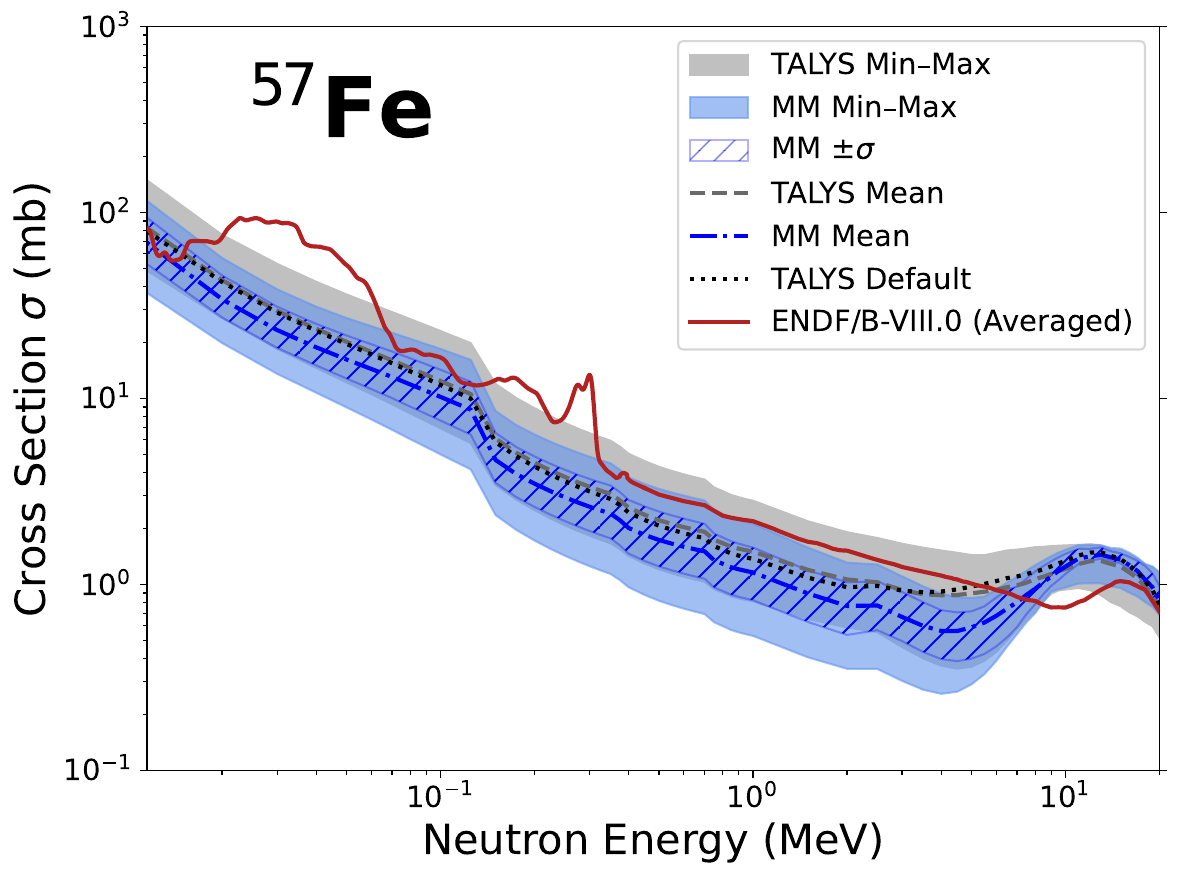}
\caption[]{Neutron capture cross section for $^{57}$Fe$(n,\gamma)$ as a function of neutron energy, 
calculated with TALYS. The gray band shows the range (min-max) obtained from all combinations 
of NLD and $\gamma$SF models available in TALYS, while the blue band corresponds 
to the full spread obtained using the MM level densities combined with all $\gamma$SF models. 
The hatched region represents the central 68\% interval of the MM-based results. The dashed and 
dash-dotted lines denote the TALYS and MM mean values, respectively, while the dotted line shows 
the TALYS default calculation. Evaluated data from the ENDF/B-VIII.0 library \cite{Brown2018} are 
shown for comparison (red solid line).}\label{fig4}
\end{figure}

\subsection{Neutron capture cross sections}

Figs. \ref{fig3} and \ref{fig4} show the calculated neutron capture cross sections 
for $^{56}$Fe$(n,\gamma)$ and $^{57}$Fe$(n,\gamma)$, respectively. The gray bands represent the full 
spread of TALYS predictions obtained by combining all available NLD and $\gamma$SF models, while the 
blue bands correspond to the full spread of results obtained using the MM level densities combined with the same set of 
$\gamma$SF models. The hatched regions indicate the central 68\% interval of the MM-based results. In both cases, the MM-based results reproduce the overall energy dependence of the 
evaluated ENDF/B-VIII.0 cross sections. The central 68\% interval is systematically narrower than the 
full TALYS model spread, although still quite wide. In Section~\ref{sec:D} we discuss 
the factors that contribute to the width of the MM spread. 

For $^{56}$Fe, the MM mean values closely follow the evaluated data over a broad energy range, with 
deviations that remain within the combined model uncertainties. For $^{57}$Fe, a similar level of 
agreement is observed, although the ENDF/B-VIII.0 data points lie consistently above the MM band. 
Resonance cross sections for low neutron energies have been averaged in 100 keV energy bins. Overall, 
the MM results lie within the range of TALYS predictions and provide a consistent description of the 
available evaluated data. 

\subsection{Maxwellian averaged cross sections}

\begin{figure}
\centering
\includegraphics[height=60mm]{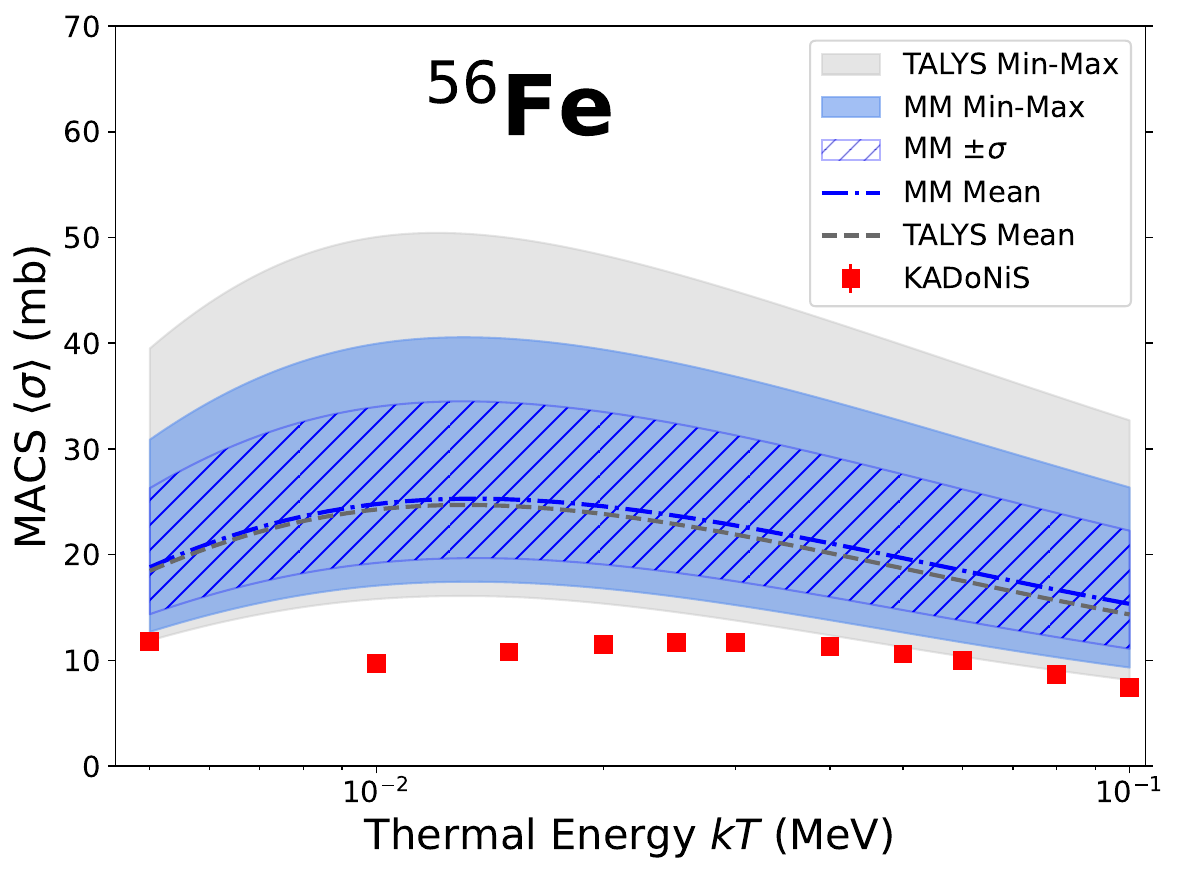}
\caption[]{Maxwellian averaged cross section (MACS) for $^{56}$Fe$(n,\gamma)$ as a function of thermal 
energy $kT$, calculated with TALYS. The gray band represents the range (min-max) obtained 
from all combinations of NLD and $\gamma$SF
models available in TALYS, while the blue band corresponds to results obtained using the 
MM level densities combined with all $\gamma$SF models. The hatched region indicates 
the central 68\% interval of the MM-based results, and the 
blue line shows the corresponding mean value. Experimental MACS values from the KADoNiS \cite{Kadonis} 
database are shown for comparison.}\label{fig5}
\end{figure}

\begin{figure}
\centering
\includegraphics[height=60mm]{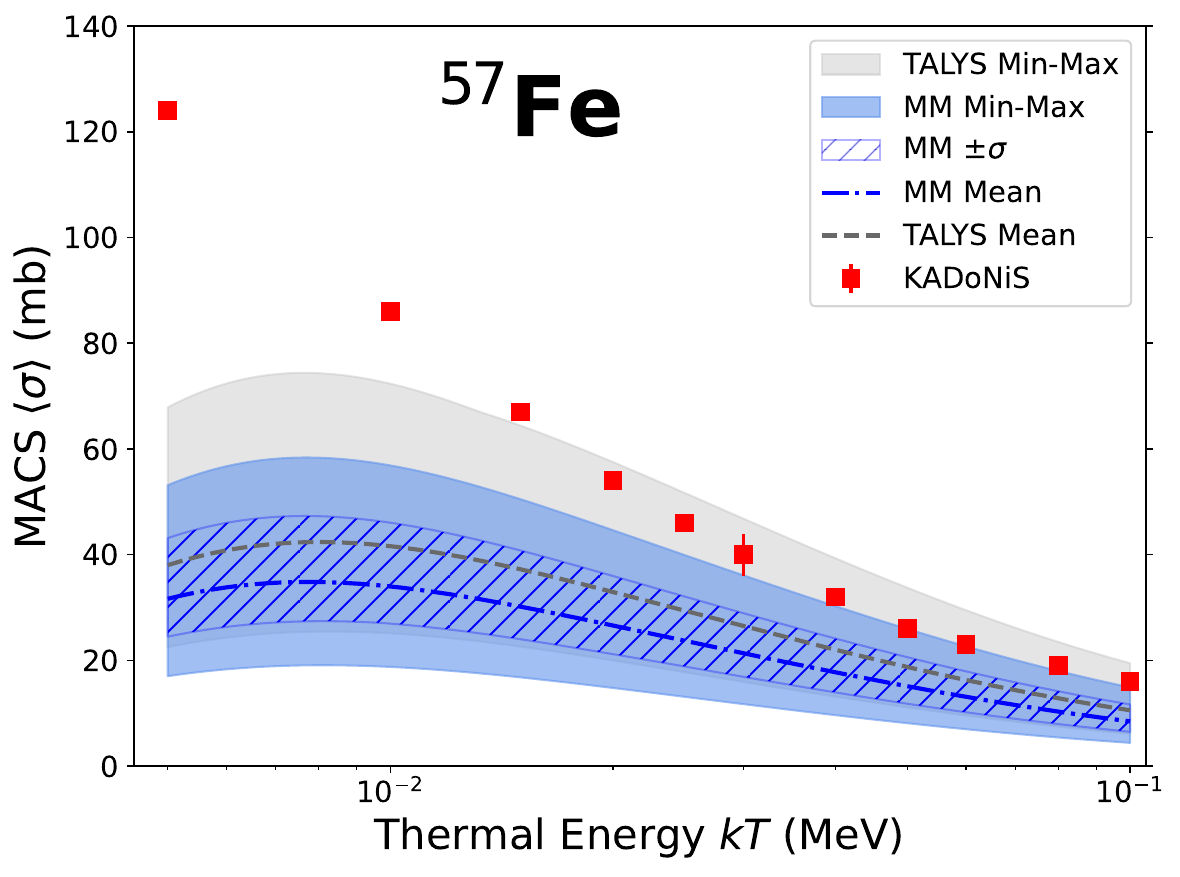}
\caption[]{Maxwellian averaged cross section (MACS) for $^{57}$Fe$(n,\gamma)$ as a function of thermal 
energy $kT$, calculated with TALYS. The gray band represents the range (min-max) obtained 
from all combinations of NLD and $\gamma$SF
models available in TALYS, while the blue band corresponds to results obtained using the 
MM level densities combined with all $\gamma$SF models. The hatched region indicates 
the central 68\% interval of the MM-based results, and the 
blue line shows the corresponding mean value. Experimental MACS values from the KADoNiS \cite{Kadonis} 
database are shown for comparison.}\label{fig6}
\end{figure}

In order to compare with KADoNiS data, we estimated the Maxwellian averaged cross sections (MACS) 
for the $^{56}$Fe$(n,\gamma)$ and $^{57}$Fe$(n,\gamma)$ reactions over a range of thermal energies 
$k_B T$ which represents the Maxwell-Boltzmann neutron energy distribution in stellar environments.
The MACS values were calculated using TALYS by averaging the energy-dependent cross sections over 
a Maxwell-Boltzmann neutron energy distribution, following the equation: 

\begin{equation}
\sigma_{MACS}\left( k_B T\right) = \frac{2}{\sqrt \pi}\frac{1}{\left(k_B T \right)^2} \int_{0}^{\infty} \sigma(E_n) E_n e^{-\frac{E_n}{k_B T}} \, dE_n
\label{eq:MACS}
\end{equation}
where $\sigma(E_n)$ is the energy dependent capture cross section.

The results are shown in Figs.~\ref{fig5} and \ref{fig6}. The gray bands represent 
the full range of TALYS predictions obtained from all combinations of NLD and $\gamma$SF models, while 
the blue bands correspond to calculations performed using the MM level densities combined with all 
available $\gamma$SF models. 

Overall, the MM-based MACS values are in good agreement with the KADoNiS data across the range of 
thermal energies considered. The central 68\% interval of the MM-based results is narrower than the full TALYS model spread, 
indicating that the use of MM level densities reduces the model dependence associated with the NLD 
input. The remaining spread is primarily driven by variations in the $\gamma$SF, 
highlighting the dominant role of $\gamma$-ray transmission coefficients in determining the reaction 
rates.

\subsection{Maxwellian averaged reaction rates}

\begin{figure}
\centering
\includegraphics[height=60mm]{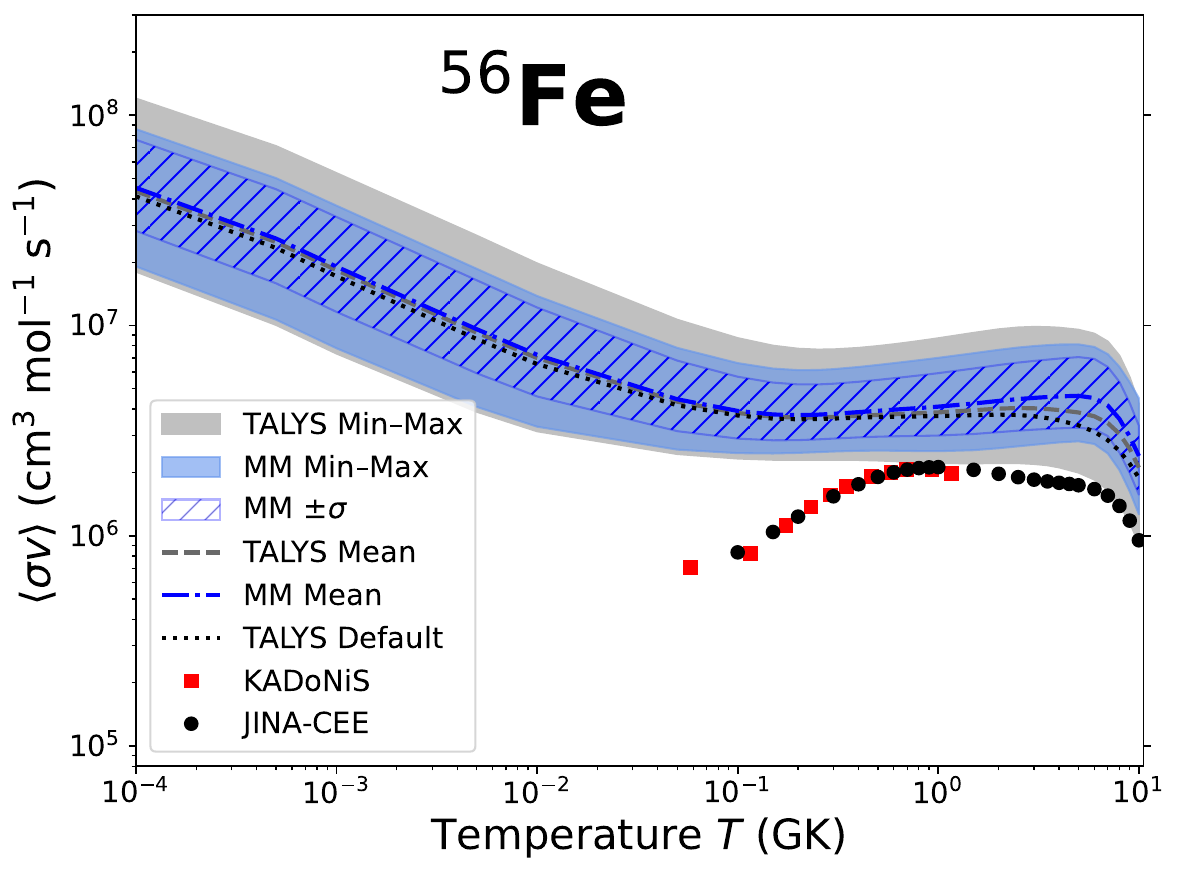}
\caption[]{Maxwellian averaged reaction rate $\langle \sigma v \rangle$ for $^{56}$Fe$(n,\gamma)$ as a function 
of temperature $T$, calculated with TALYS. The gray band represents the range (min-max) obtained 
from all combinations of NLD and $\gamma$SF models 
available in TALYS, while the blue band corresponds to results obtained using the MM 
level densities combined with all $\gamma$SF models. The hatched region indicates the central 68\% interval of the MM-based results, and the 
blue dash-dotted line shows the corresponding mean value. The dashed and dotted lines represent the TALYS mean and 
default calculations, respectively. Recommended reaction rates from the KADoNiS \cite{Kadonis} and JINA-CEE \cite{JINA} 
databases are shown for comparison.}\label{fig7}
\end{figure}

\begin{figure}
\centering
\includegraphics[height=60mm]{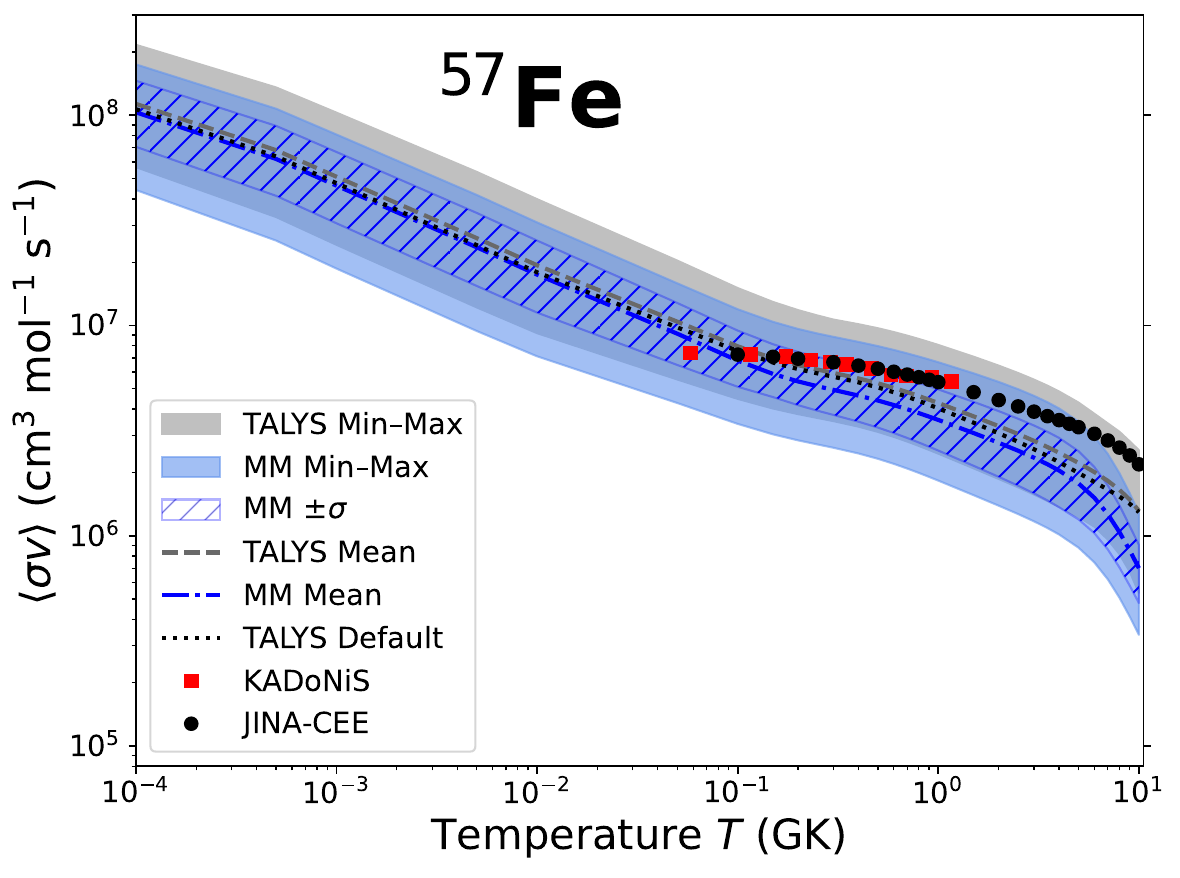}
\caption[]{Maxwellian averaged reaction rate $\langle \sigma v \rangle$ for $^{57}$Fe$(n,\gamma)$ as a function 
of temperature $T$, calculated with TALYS. The gray band represents the range (min-max) obtained 
from all combinations of NLD and $\gamma$SF models 
available in TALYS, while the blue band corresponds to results obtained using the MM 
level densities combined with all $\gamma$SF models. The hatched region indicates the central 68\% interval of the MM-based results, and the 
blue dash-dotted line shows the corresponding mean value. The dashed and dotted lines represent the TALYS mean and 
default calculations, respectively. Recommended reaction rates from the KADoNiS \cite{Kadonis} and JINA-CEE \cite{JINA} 
databases are shown for comparison.}\label{fig8}
\end{figure}

The Maxwellian averaged reaction rates for the $^{56}$Fe$(n,\gamma)$ and $^{57}$Fe$(n,\gamma)$ 
reactions are shown in Figs.~\ref{fig7} and \ref{fig8}, respectively, as a function 
of temperature. The gray bands represent the full range of TALYS predictions obtained from all 
combinations of NLD and $\gamma$SF models, while the blue bands correspond to results obtained using the
MM level densities combined with the same set of $\gamma$SF models. The hatched regions indicate 
the central 68\% interval of the MM-based results.

In both cases, the reaction rates exhibit a smooth decrease with increasing temperature, reflecting 
the underlying energy dependence of the neutron capture cross sections. The MM-based results follow 
the overall trend of the TALYS calculations and lie within the full model spread across the entire 
temperature range considered. At low temperatures, corresponding to astrophysically relevant 
conditions, the central 68\% interval of the MM-based results is significantly narrower than the full TALYS spread, indicating 
that the use of MM level densities reduces the model dependence associated with the NLD input.

The comparison with recommended reaction rates from the KADoNiS and JINA-CEE databases shows that 
the MM results are in good agreement with the available data for both nuclei. For $^{57}$Fe, the 
agreement is particularly good over the temperature range where experimental constraints are available. 
For $^{56}$Fe, somewhat larger deviations are observed at intermediate temperatures, although the 
MM results remain within the combined model uncertainties. The remaining spread in the calculated 
reaction rates is primarily driven by variations in the $\gamma$SF, highlighting the dominant role of 
$\gamma$-ray transmission coefficients in the Hauser-Feshbach description of neutron capture reactions. 
Overall, the MM level densities provide a consistent description of the reaction rates.


\section{\label{sec:D}Discussion of results}

\begin{figure}
\centering
\includegraphics[height=60mm]{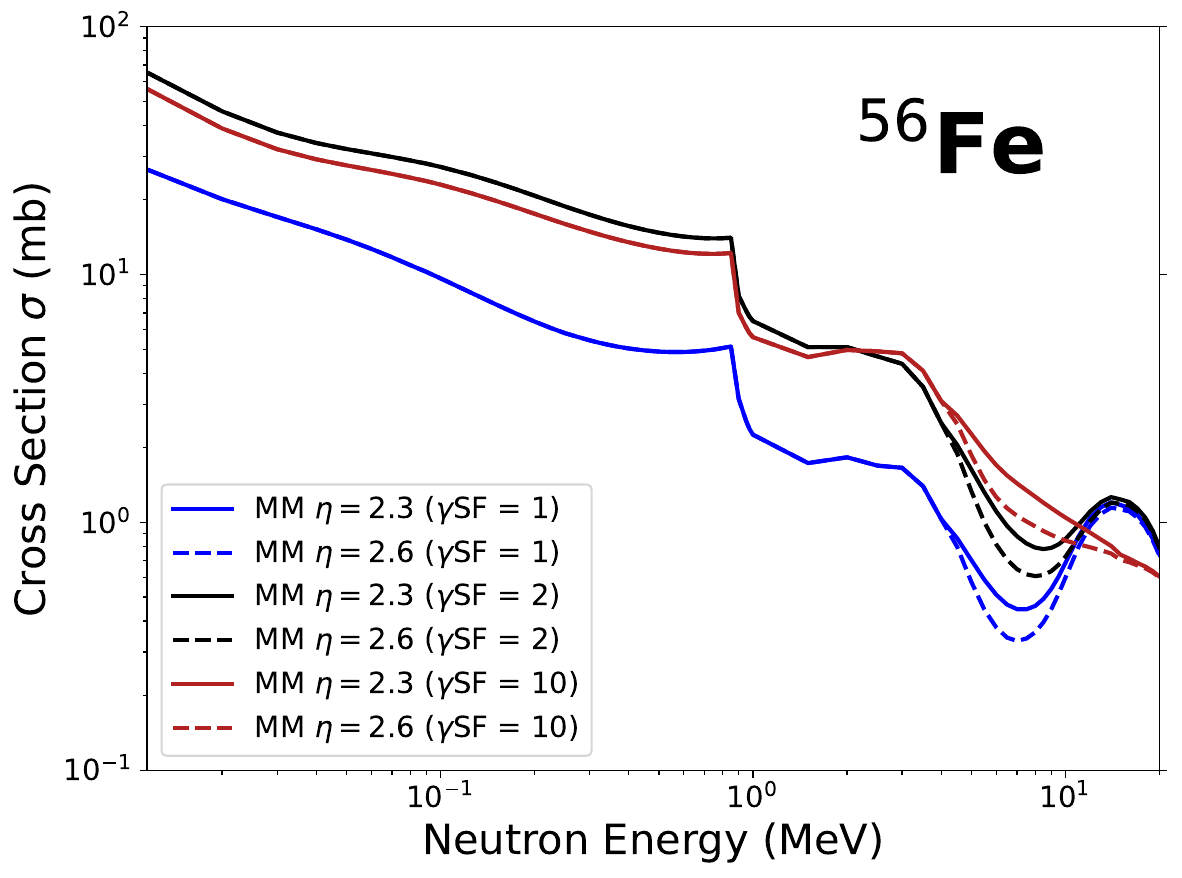}
\caption[]{Neutron capture cross section for $^{56}$Fe$(n,\gamma)$ as a function of neutron energy, 
calculated using the MM level densities for two values of the cutoff parameter, 
$\eta=2.3$ and $\eta=2.6$, combined with selected $\gamma$SF models available in TALYS. 
The $\gamma$SF models considered are $\gamma$SF=1, $\gamma$SF=2, and $\gamma$SF=10, with 
$\gamma$SF=1, $\gamma$SF=2 corresponding to different Lorentzian parameterizations of the E1 strength 
and $\gamma$SF=10 to Skyrme-BSK27 HFB+QRPA tables. The $\gamma$SF=1 model yields the lowest cross 
sections, and $\gamma$SF=2 and 10 the highest. The variation between $\eta$ values reflects the 
variation associated with the MM level densities, while the spread between $\gamma$SF models 
illustrates the dominant sensitivity of the cross section to the $\gamma$SF.}\label{fig9}
\end{figure}

\begin{figure}
\centering
\includegraphics[height=60mm]{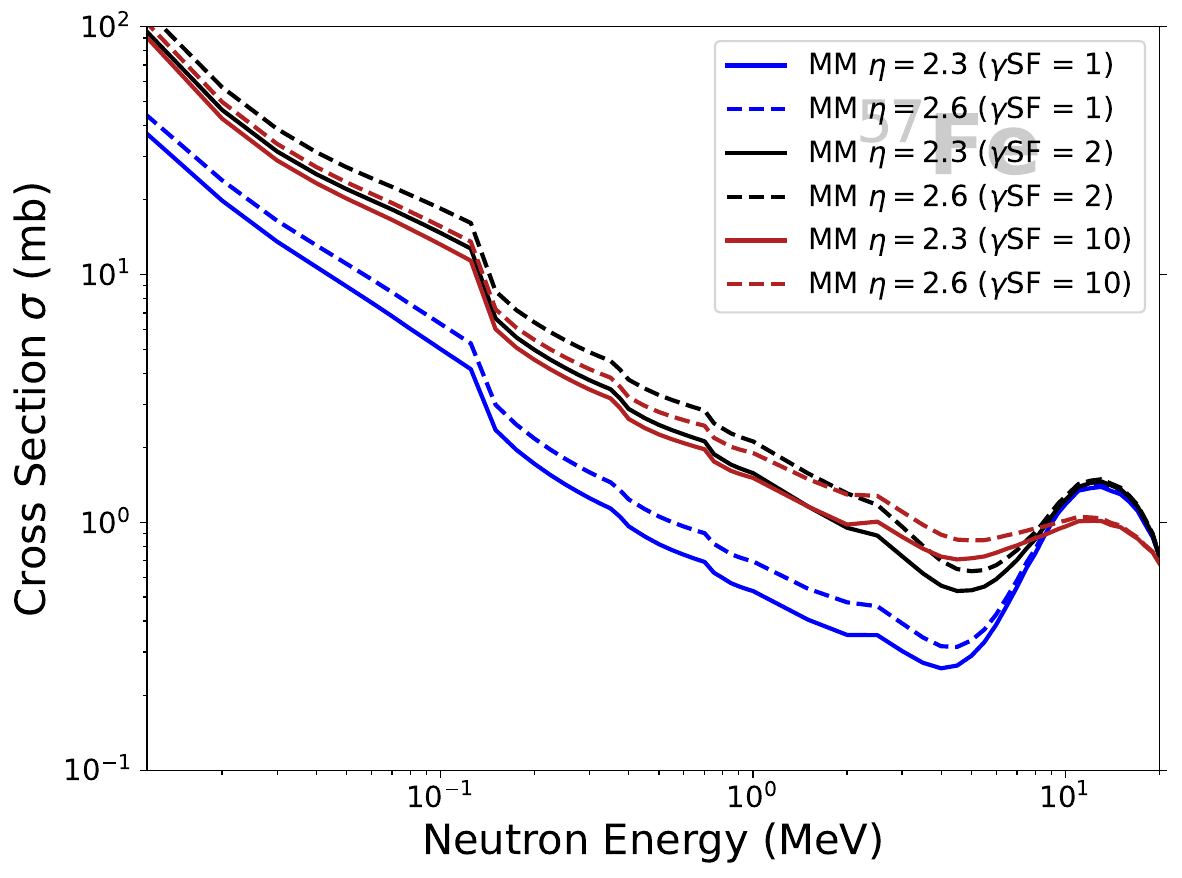}
\caption[]{Neutron capture cross section for $^{57}$Fe$(n,\gamma)$ as a function of neutron energy, 
calculated using the MM level densities for two values of the cutoff parameter, 
$\eta=2.3$ and $\eta=2.6$, combined with selected $\gamma$SF models available in TALYS. 
The $\gamma$SF models considered are $\gamma$SF=1, $\gamma$SF=2, and $\gamma$SF=10, with 
$\gamma$SF=1, $\gamma$SF=2 corresponding to different Lorentzian parameterizations of the E1 strength 
and $\gamma$SF=10 to Skyrme-BSK27 HFB+QRPA tables. The $\gamma$SF=1 model yields the lowest cross 
sections, and $\gamma$SF=2 and 10 the highest. The variation between $\eta$ values reflects the 
variation associated with the MM level densities, while the spread between $\gamma$SF models 
illustrates the dominant sensitivity of the cross section to the $\gamma$SF.}\label{fig10}
\end{figure}

The results show that the calculated cross sections exhibit a noticeable dependence on both the NLD and 
the $\gamma$SF. In particular, variations in the $\gamma$SF 
lead to significant differences in the magnitude of the cross sections, reflecting the dominant role 
of the $\gamma$-ray transmission coefficients in the Hauser-Feshbach framework. This behavior is clearly 
illustrated in Figs.~\ref{fig9} and \ref{fig10}, where cross sections calculated using the MM level 
densities for two values of the cutoff parameter $\eta$ are combined with selected $\gamma$SF models. 
In both nuclei, the GS=1 model yields the lowest cross sections, GS=2 the highest, and GS=9 intermediate 
values, demonstrating a systematic ordering governed by the choice of $\gamma$SF. The variation 
associated with the NLD, as reflected by the two $\eta$ values, is comparatively minimal.

\begin{figure}
\centering
\includegraphics[height=60mm]{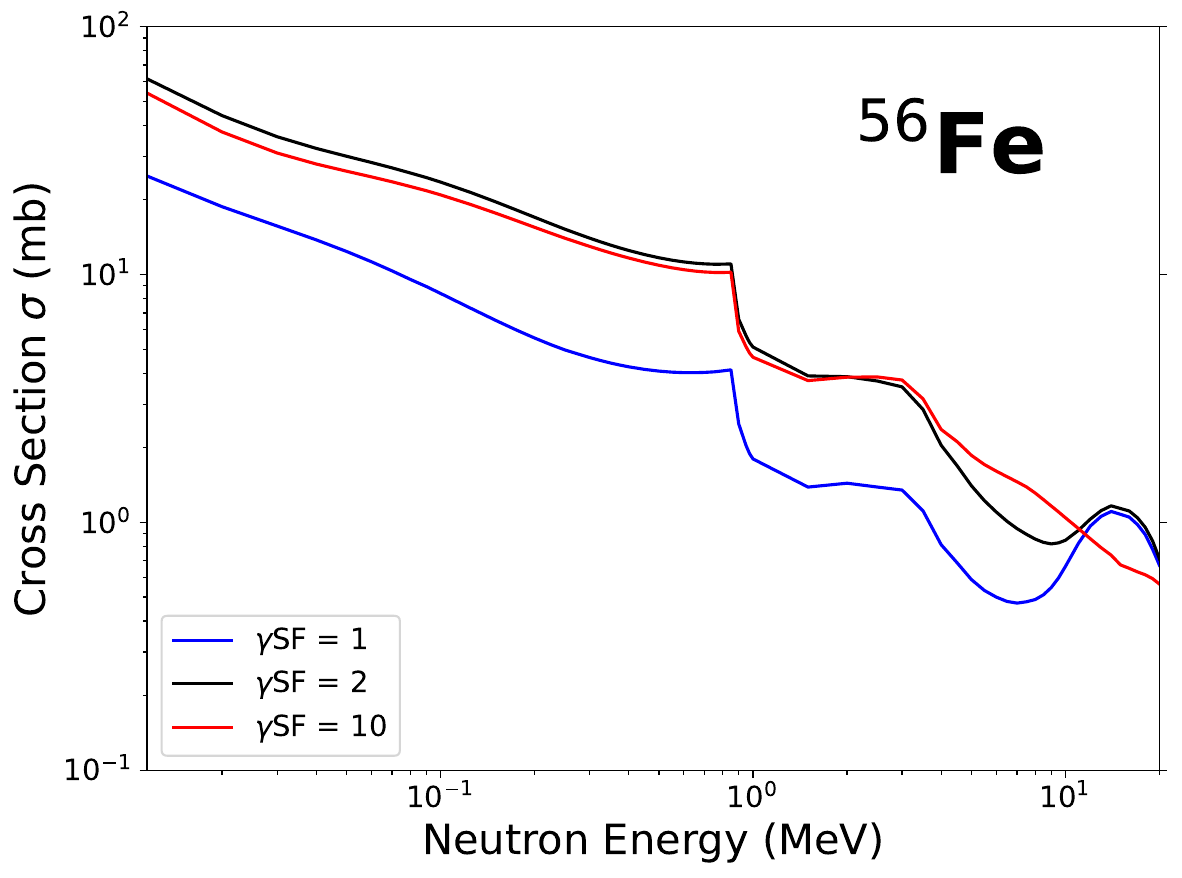}
\caption[]{Neutron capture cross section for $^{56}$Fe$(n,\gamma)$  as a function of neutron energy, 
calculated with TALYS using a fixed nuclear level density model (NLD = 1, constant temperature model) 
and varying $\gamma$SF models ($\gamma$SF = 1, 2, 10).}\label{fig11}
\end{figure}

\begin{figure}
\centering
\includegraphics[height=60mm]{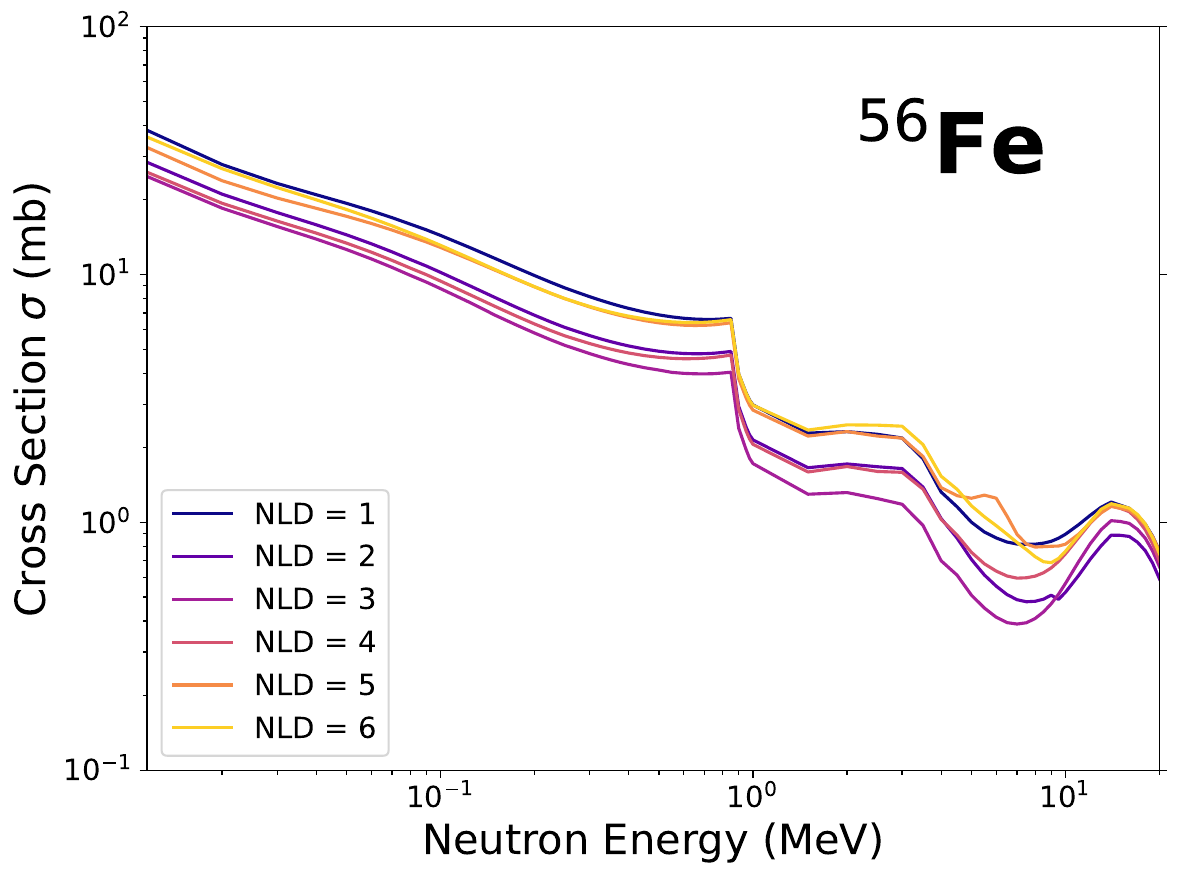}
\caption[]{Neutron capture cross section for $^{56}$Fe$(n,\gamma)$  as a function of neutron energy, 
calculated with TALYS using a fixed $\gamma$SF ($\gamma$SF = 9) and varying nuclear level density 
models (NLD = 1--6).}\label{fig12}
\end{figure}

This hierarchy of sensitivities is further emphasized in Fig.~\ref{fig11}, where the NLD is fixed 
(LD = 1) and all available $\gamma$SF models in TALYS are considered. The resulting spread in the 
cross sections spans more than an order of magnitude over parts of the energy range, highlighting the 
dominant impact of the $\gamma$SF on the calculated observables. In contrast, Fig.~\ref{fig12} shows 
the variation obtained by fixing the $\gamma$SF (GS = 9) and varying the NLD models (LD = 1--6). In 
this case, the spread is significantly smaller, indicating that the sensitivity to the NLD is secondary 
compared to that of the $\gamma$SF, although differences between phenomenological and microscopic level 
densities remain visible.

Comparisons using fixed $\gamma$SFs and different NLD models indicate that the MM level densities produce 
cross sections that are generally consistent with those obtained using microscopic HFB-based models, while 
also providing a sensitivity band through the variation of the cutoff parameter $\eta$. 

The Maxwellian averaged reaction rates display a similar sensitivity to the underlying nuclear inputs. 
The inclusion of experimental constraints, when available, reduces the spread of the calculated rates; 
however, significant variations remain depending on the choice of $\gamma$SF. The MM-based results 
typically fall within the range defined by the TALYS models and are in reasonable agreement with the 
recommended values from KADoNiS and JINA.

Additional calculations of Maxwellian averaged cross sections (MACS) were performed and compared directly 
with KADoNiS data. The MM-based MACS values are generally consistent with the experimental trends, with 
deviations that reflect the combined impact of uncertainties in both the NLD and $\gamma$SF inputs.

Overall, the results demonstrate that the choice of $\gamma$SF constitutes the dominant source 
of uncertainty in the calculated neutron-capture observables, while the nuclear level density plays a 
secondary but non-negligible role. The MM level densities provide a viable microscopic alternative for use 
in Hauser-Feshbach calculations, offering a consistent description of the NLD input together with a 
transparent sensitivity estimate associated with the MM cutoff-parameter variation.

\section{\label{sec:C}Conclusion}

In this work, we investigated the impact of nuclear level densities (NLDs) obtained with the 
Moments Method (MM) on $(n,\gamma)$ cross sections and Maxwellian-averaged reaction rates for 
$^{56}$Fe and $^{57}$Fe within the Hauser-Feshbach framework using the TALYS reaction code. 
The MM level densities were implemented in TALYS and combined with a range of $\gamma$-ray 
strength function ($\gamma$SF) models, enabling a systematic study of the sensitivity of 
neutron capture observables to nuclear statistical properties.

The results show that the calculated cross sections and reaction rates exhibit a strong dependence 
on the choice of $\gamma$SF model, while the sensitivity to the NLD is comparatively weaker, but 
nevertheless still important. This hierarchy is clearly demonstrated by calculations in which either 
the $\gamma$SF or the NLD is held fixed, revealing that variations in the $\gamma$SF lead to 
significantly larger changes in the predicted observables. This behavior reflects the dominant role 
of the $\gamma$-ray transmission coefficients in the statistical model description of neutron-capture 
reactions.

Within the MM framework, uncertainties in the NLD are quantified through the variation of the cutoff 
parameter $\eta$ in the range $2.3 \leq \eta \leq 2.6$. In the present work this provides a practical sensitivity estimate associated with the MM input, which propagates to the calculated cross 
sections and reaction rates. The resulting MM sensitivity bands are consistently narrower than the 
full spread of TALYS predictions obtained using different NLD models.

The calculated cross sections are in reasonable agreement with evaluated data from the ENDF/B-VIII.0 
library, while the Maxwellian averaged cross sections and reaction rates show good agreement with 
recommended values from the KADoNiS and JINA-CEE databases. The remaining discrepancies are primarily 
associated with uncertainties in the $\gamma$SF, highlighting the need for improved constraints on the 
electromagnetic response of nuclei.

Overall, the present study demonstrates that the Moments Method provides a viable microscopic 
alternative for nuclear level densities in Hauser-Feshbach calculations, within a transparent and 
reproducible sensitivity-study framework. The results further indicate that reducing uncertainties in 
$\gamma$-ray strength functions is essential for improving the predictive power of neutron capture 
calculations, particularly for applications to nuclei far from stability. A more rigorous treatment 
of these uncertainties, for example within a Bayesian framework, will be pursued in future work.

\section*{Data Availability}
The data that support the findings of this article are available from the authors upon reasonable request.

\begin{acknowledgments}
Useful discussion with C. Larsen are gratefully acknowledged. This material is based upon work supported by the National Science Foundation under Grants No. PHY--2412851.
\end{acknowledgments}

\bibliography{Fe56_57_LD8}

\providecommand{\noopsort}[1]{}\providecommand{\singleletter}[1]{#1}%
\begin{thebibliography}{71}%
\makeatletter
\providecommand \@ifxundefined [1]{%
 \@ifx{#1\undefined}
}%
\providecommand \@ifnum [1]{%
 \ifnum #1\expandafter \@firstoftwo
 \else \expandafter \@secondoftwo
 \fi
}%
\providecommand \@ifx [1]{%
 \ifx #1\expandafter \@firstoftwo
 \else \expandafter \@secondoftwo
 \fi
}%
\providecommand \natexlab [1]{#1}%
\providecommand \enquote  [1]{``#1''}%
\providecommand \bibnamefont  [1]{#1}%
\providecommand \bibfnamefont [1]{#1}%
\providecommand \citenamefont [1]{#1}%
\providecommand \href@noop [0]{\@secondoftwo}%
\providecommand \href [0]{\begingroup \@sanitize@url \@href}%
\providecommand \@href[1]{\@@startlink{#1}\@@href}%
\providecommand \@@href[1]{\endgroup#1\@@endlink}%
\providecommand \@sanitize@url [0]{\catcode `\\12\catcode `\$12\catcode
  `\&12\catcode `\#12\catcode `\^12\catcode `\_12\catcode `\%12\relax}%
\providecommand \@@startlink[1]{}%
\providecommand \@@endlink[0]{}%
\providecommand \url  [0]{\begingroup\@sanitize@url \@url }%
\providecommand \@url [1]{\endgroup\@href {#1}{\urlprefix }}%
\providecommand \urlprefix  [0]{URL }%
\providecommand \Eprint [0]{\href }%
\providecommand \doibase [0]{https://doi.org/}%
\providecommand \selectlanguage [0]{\@gobble}%
\providecommand \bibinfo  [0]{\@secondoftwo}%
\providecommand \bibfield  [0]{\@secondoftwo}%
\providecommand \translation [1]{[#1]}%
\providecommand \BibitemOpen [0]{}%
\providecommand \bibitemStop [0]{}%
\providecommand \bibitemNoStop [0]{.\EOS\space}%
\providecommand \EOS [0]{\spacefactor3000\relax}%
\providecommand \BibitemShut  [1]{\csname bibitem#1\endcsname}%
\let\auto@bib@innerbib\@empty
\bibitem [{\citenamefont {Pignatari}\ \emph {et~al.}(2010)\citenamefont
  {Pignatari}, \citenamefont {Gallino}, \citenamefont {Heil}, \citenamefont
  {Wiescher}, \citenamefont {Käppeler}, \citenamefont {Herwig},\ and\
  \citenamefont {Bisterzo}}]{Pignatari_2010}%
  \BibitemOpen
  \bibfield  {author} {\bibinfo {author} {\bibfnamefont {M.}~\bibnamefont
  {Pignatari}}, \bibinfo {author} {\bibfnamefont {R.}~\bibnamefont {Gallino}},
  \bibinfo {author} {\bibfnamefont {M.}~\bibnamefont {Heil}}, \bibinfo {author}
  {\bibfnamefont {M.}~\bibnamefont {Wiescher}}, \bibinfo {author}
  {\bibfnamefont {F.}~\bibnamefont {Käppeler}}, \bibinfo {author}
  {\bibfnamefont {F.}~\bibnamefont {Herwig}},\ and\ \bibinfo {author}
  {\bibfnamefont {S.}~\bibnamefont {Bisterzo}},\ }\bibfield  {title} {\bibinfo
  {title} {The weak s-process in massive stars and its dependence on the
  neutron capture cross sections},\ }\href
  {https://doi.org/10.1088/0004-637X/710/2/1557} {\bibfield  {journal}
  {\bibinfo  {journal} {The Astrophysical Journal}\ }\textbf {\bibinfo {volume}
  {710}},\ \bibinfo {pages} {1557} (\bibinfo {year} {2010})}\BibitemShut
  {NoStop}%
\bibitem [{\citenamefont {Käppeler}\ \emph {et~al.}(2011)\citenamefont
  {Käppeler}, \citenamefont {Gallino}, , \citenamefont {Bisterzo},\ and\
  \citenamefont {Aoki}}]{RevModPhys.83.157}%
  \BibitemOpen
  \bibfield  {author} {\bibinfo {author} {\bibfnamefont {F.}~\bibnamefont
  {Käppeler}}, \bibinfo {author} {\bibfnamefont {R.}~\bibnamefont {Gallino}},
  , \bibinfo {author} {\bibfnamefont {S.}~\bibnamefont {Bisterzo}},\ and\
  \bibinfo {author} {\bibfnamefont {W.}~\bibnamefont {Aoki}},\ }\bibfield
  {title} {\bibinfo {title} {The $s$ process: Nuclear physics, stellar models,
  and observations},\ }\href {https://doi.org/10.1103/RevModPhys.83.157}
  {\bibfield  {journal} {\bibinfo  {journal} {Rev. Mod. Phys.}\ }\textbf
  {\bibinfo {volume} {83}},\ \bibinfo {pages} {157} (\bibinfo {year}
  {2011})}\BibitemShut {NoStop}%
\bibitem [{\citenamefont {Arnould}\ \emph {et~al.}(2007)\citenamefont
  {Arnould}, \citenamefont {Goriely},\ and\ \citenamefont
  {Takahashi}}]{ARNOULD200797}%
  \BibitemOpen
  \bibfield  {author} {\bibinfo {author} {\bibfnamefont {M.}~\bibnamefont
  {Arnould}}, \bibinfo {author} {\bibfnamefont {S.}~\bibnamefont {Goriely}},\
  and\ \bibinfo {author} {\bibfnamefont {K.}~\bibnamefont {Takahashi}},\ }\href
  {https://doi.org/https://doi.org/10.1016/j.physrep.2007.06.002} {\bibfield
  {journal} {\bibinfo  {journal} {Physics Reports}\ }\textbf {\bibinfo {volume}
  {450}},\ \bibinfo {pages} {97} (\bibinfo {year} {2007})}\BibitemShut
  {NoStop}%
\bibitem [{\citenamefont {Cowan}\ and\ \citenamefont
  {Rose}(1977)}]{Cowan_1977}%
  \BibitemOpen
  \bibfield  {author} {\bibinfo {author} {\bibfnamefont {J.}~\bibnamefont
  {Cowan}}\ and\ \bibinfo {author} {\bibfnamefont {W.}~\bibnamefont {Rose}},\
  }\href {https://doi.org/10.1086/155030} {\bibfield  {journal} {\bibinfo
  {journal} {Astrophys. J.}\ }\textbf {\bibinfo {volume} {212}},\ \bibinfo
  {pages} {149} (\bibinfo {year} {1977})}\BibitemShut {NoStop}%
\bibitem [{\citenamefont {Sneden}\ \emph {et~al.}(2008)\citenamefont {Sneden},
  \citenamefont {Cowan},\ and\ \citenamefont {Gallino}}]{Sneden_2008}%
  \BibitemOpen
  \bibfield  {author} {\bibinfo {author} {\bibfnamefont {C.}~\bibnamefont
  {Sneden}}, \bibinfo {author} {\bibfnamefont {J.}~\bibnamefont {Cowan}},\ and\
  \bibinfo {author} {\bibfnamefont {R.}~\bibnamefont {Gallino}},\ }\href
  {https://doi.org/10.1146/annurev.astro.46.060407.145207} {\bibfield
  {journal} {\bibinfo  {journal} {Annu. Rev. of Astron. Astrophys.}\ }\textbf
  {\bibinfo {volume} {46}},\ \bibinfo {pages} {241} (\bibinfo {year}
  {2008})}\BibitemShut {NoStop}%
\bibitem [{\citenamefont {Gelles}(1990)}]{Gelles_1990}%
  \BibitemOpen
  \bibfield  {author} {\bibinfo {author} {\bibfnamefont {D.~S.}\ \bibnamefont
  {Gelles}},\ }\href@noop {} {\bibfield  {journal} {\bibinfo  {journal} {ISIJ
  International (Iron and Steel Institute of Japan)}\ }\textbf {\bibinfo
  {volume} {30}},\ \bibinfo {pages} {905} (\bibinfo {year} {1990})}\BibitemShut
  {NoStop}%
\bibitem [{\citenamefont {Terrani}\ \emph {et~al.}(2014)\citenamefont
  {Terrani}, \citenamefont {Zinkle},\ and\ \citenamefont
  {Snead}}]{TERRANI2014420}%
  \BibitemOpen
  \bibfield  {author} {\bibinfo {author} {\bibfnamefont {K.}~\bibnamefont
  {Terrani}}, \bibinfo {author} {\bibfnamefont {S.}~\bibnamefont {Zinkle}},\
  and\ \bibinfo {author} {\bibfnamefont {L.}~\bibnamefont {Snead}},\ }\href
  {https://doi.org/https://doi.org/10.1016/j.jnucmat.2013.06.041} {\bibfield
  {journal} {\bibinfo  {journal} {Journal of Nuclear Materials}\ }\textbf
  {\bibinfo {volume} {448}},\ \bibinfo {pages} {420} (\bibinfo {year}
  {2014})}\BibitemShut {NoStop}%
\bibitem [{\citenamefont {Rebak}(2017)}]{Rebak_2017}%
  \BibitemOpen
  \bibfield  {author} {\bibinfo {author} {\bibfnamefont {R.}~\bibnamefont
  {Rebak}},\ }\href {https://doi.org/10.1051/epjn/2017029} {\bibfield
  {journal} {\bibinfo  {journal} {EPJ Nuclear Sci. Technol.}\ }\textbf
  {\bibinfo {volume} {3}},\ \bibinfo {pages} {34} (\bibinfo {year}
  {2017})}\BibitemShut {NoStop}%
\bibitem [{\citenamefont {Colonna}\ \emph {et~al.}(2010)\citenamefont
  {Colonna}, \citenamefont {Belloni}, \citenamefont {Berthoumieux},
  \citenamefont {Calviani}, \citenamefont {Domingo-Pardo}, \citenamefont
  {Guerrero}, \citenamefont {Karadimos}, \citenamefont {Lederer}, \citenamefont
  {Massimi}, \citenamefont {Paradela}, \citenamefont {Plag}, \citenamefont
  {Praena},\ and\ \citenamefont {Sarmento}}]{Colonna_2010}%
  \BibitemOpen
  \bibfield  {author} {\bibinfo {author} {\bibfnamefont {N.}~\bibnamefont
  {Colonna}}, \bibinfo {author} {\bibfnamefont {F.}~\bibnamefont {Belloni}},
  \bibinfo {author} {\bibfnamefont {E.}~\bibnamefont {Berthoumieux}}, \bibinfo
  {author} {\bibfnamefont {M.}~\bibnamefont {Calviani}}, \bibinfo {author}
  {\bibfnamefont {C.}~\bibnamefont {Domingo-Pardo}}, \bibinfo {author}
  {\bibfnamefont {C.}~\bibnamefont {Guerrero}}, \bibinfo {author}
  {\bibfnamefont {D.}~\bibnamefont {Karadimos}}, \bibinfo {author}
  {\bibfnamefont {C.}~\bibnamefont {Lederer}}, \bibinfo {author} {\bibfnamefont
  {C.}~\bibnamefont {Massimi}}, \bibinfo {author} {\bibfnamefont
  {C.}~\bibnamefont {Paradela}}, \bibinfo {author} {\bibfnamefont
  {R.}~\bibnamefont {Plag}}, \bibinfo {author} {\bibfnamefont {J.}~\bibnamefont
  {Praena}},\ and\ \bibinfo {author} {\bibfnamefont {R.}~\bibnamefont
  {Sarmento}},\ }\href {https://doi.org/10.1039/C0EE00108B} {\bibfield
  {journal} {\bibinfo  {journal} {Energy Environ. Sci.}\ }\textbf {\bibinfo
  {volume} {3}},\ \bibinfo {pages} {1910} (\bibinfo {year} {2010})}\BibitemShut
  {NoStop}%
\bibitem [{NEA(2002)}]{NEA_2002}%
  \BibitemOpen
  \href@noop {} {\emph {\bibinfo {title} {NEA, Accelerator-driven Systems (ADS)
  and Fast Reactors (FR) in Advanced Nuclear Fuel Cylces}}}\ (\bibinfo
  {publisher} {OECD Publishing},\ \bibinfo {address} {Paris},\ \bibinfo {year}
  {2002})\BibitemShut {NoStop}%
\bibitem [{\citenamefont {Knapov\'a}\ \emph {et~al.}(2025)\citenamefont
  {Knapov\'a}, \citenamefont {Hor\ifmmode \check{c}\else
  \v{c}\fi{}i\ifmmode~\check{c}\else \v{c}\fi{}kov\'a}, \citenamefont
  {Couture}, \citenamefont {Fry}, \citenamefont {Gunsing}, \citenamefont
  {Kawano}, \citenamefont {Kelly}, \citenamefont {Krti\ifmmode~\check{c}\else
  \v{c}\fi{}ka}, \citenamefont {Cidoncha}, \citenamefont {Prokop},
  \citenamefont {Reifarth}, \citenamefont {Rusev}, \citenamefont {Ullmann},\
  and\ \citenamefont {Valenta}}]{Knapova_25}%
  \BibitemOpen
  \bibfield  {author} {\bibinfo {author} {\bibfnamefont {I.}~\bibnamefont
  {Knapov\'a}}, \bibinfo {author} {\bibfnamefont {K.}~\bibnamefont {Hor\ifmmode
  \check{c}\else \v{c}\fi{}i\ifmmode~\check{c}\else \v{c}\fi{}kov\'a}},
  \bibinfo {author} {\bibfnamefont {A.}~\bibnamefont {Couture}}, \bibinfo
  {author} {\bibfnamefont {C.}~\bibnamefont {Fry}}, \bibinfo {author}
  {\bibfnamefont {F.}~\bibnamefont {Gunsing}}, \bibinfo {author} {\bibfnamefont
  {T.}~\bibnamefont {Kawano}}, \bibinfo {author} {\bibfnamefont {K.~J.}\
  \bibnamefont {Kelly}}, \bibinfo {author} {\bibfnamefont {M.}~\bibnamefont
  {Krti\ifmmode~\check{c}\else \v{c}\fi{}ka}}, \bibinfo {author} {\bibfnamefont
  {E.~L.}\ \bibnamefont {Cidoncha}}, \bibinfo {author} {\bibfnamefont {C.~J.}\
  \bibnamefont {Prokop}}, \bibinfo {author} {\bibfnamefont {R.}~\bibnamefont
  {Reifarth}}, \bibinfo {author} {\bibfnamefont {G.}~\bibnamefont {Rusev}},
  \bibinfo {author} {\bibfnamefont {J.~L.}\ \bibnamefont {Ullmann}},\ and\
  \bibinfo {author} {\bibfnamefont {S.}~\bibnamefont {Valenta}},\ }\href
  {https://doi.org/10.1103/5sjr-vyjv} {\bibfield  {journal} {\bibinfo
  {journal} {Phys. Rev. C}\ }\textbf {\bibinfo {volume} {112}},\ \bibinfo
  {pages} {014612} (\bibinfo {year} {2025})}\BibitemShut {NoStop}%
\bibitem [{\citenamefont {Larsen}\ \emph {et~al.}(2019)\citenamefont {Larsen},
  \citenamefont {Spyrou}, \citenamefont {Liddick},\ and\ \citenamefont
  {Guttormsen}}]{Larsen_2019}%
  \BibitemOpen
  \bibfield  {author} {\bibinfo {author} {\bibfnamefont {A.}~\bibnamefont
  {Larsen}}, \bibinfo {author} {\bibfnamefont {A.}~\bibnamefont {Spyrou}},
  \bibinfo {author} {\bibfnamefont {S.}~\bibnamefont {Liddick}},\ and\ \bibinfo
  {author} {\bibfnamefont {M.}~\bibnamefont {Guttormsen}},\ }\href
  {https://doi.org/https://doi.org/10.1016/j.ppnp.2019.04.002} {\bibfield
  {journal} {\bibinfo  {journal} {Progress in Particle and Nuclear Physics}\
  }\textbf {\bibinfo {volume} {107}},\ \bibinfo {pages} {69} (\bibinfo {year}
  {2019})}\BibitemShut {NoStop}%
\bibitem [{\citenamefont {Guttormsen}\ \emph {et~al.}(1987)\citenamefont
  {Guttormsen}, \citenamefont {y},\ and\ \citenamefont
  {Rekstad}}]{Guttormsen_1987}%
  \BibitemOpen
  \bibfield  {author} {\bibinfo {author} {\bibfnamefont {M.}~\bibnamefont
  {Guttormsen}}, \bibinfo {author} {\bibfnamefont {T.~R.}\ \bibnamefont {y}},\
  and\ \bibinfo {author} {\bibfnamefont {J.}~\bibnamefont {Rekstad}},\ }\href
  {https://doi.org/https://doi.org/10.1016/0168-9002(87)91221-6} {\bibfield
  {journal} {\bibinfo  {journal} {Nuclear Instruments and Methods in Physics
  Research Section A: Accelerators, Spectrometers, Detectors and Associated
  Equipment}\ }\textbf {\bibinfo {volume} {255}},\ \bibinfo {pages} {518}
  (\bibinfo {year} {1987})}\BibitemShut {NoStop}%
\bibitem [{\citenamefont {Schiller}\ \emph {et~al.}(2000)\citenamefont
  {Schiller}, \citenamefont {Bergholt}, \citenamefont {Guttormsen},
  \citenamefont {Melby}, \citenamefont {Rekstad},\ and\ \citenamefont
  {Siem}}]{Schiller_2000}%
  \BibitemOpen
  \bibfield  {author} {\bibinfo {author} {\bibfnamefont {A.}~\bibnamefont
  {Schiller}}, \bibinfo {author} {\bibfnamefont {L.}~\bibnamefont {Bergholt}},
  \bibinfo {author} {\bibfnamefont {M.}~\bibnamefont {Guttormsen}}, \bibinfo
  {author} {\bibfnamefont {E.}~\bibnamefont {Melby}}, \bibinfo {author}
  {\bibfnamefont {J.}~\bibnamefont {Rekstad}},\ and\ \bibinfo {author}
  {\bibfnamefont {S.}~\bibnamefont {Siem}},\ }\href
  {https://doi.org/https://doi.org/10.1016/S0168-9002(99)01187-0} {\bibfield
  {journal} {\bibinfo  {journal} {Nuclear Instruments and Methods in Physics
  Research Section A: Accelerators, Spectrometers, Detectors and Associated
  Equipment}\ }\textbf {\bibinfo {volume} {447}},\ \bibinfo {pages} {498}
  (\bibinfo {year} {2000})}\BibitemShut {NoStop}%
\bibitem [{\citenamefont {Ingeberg}\ \emph {et~al.}(2000)\citenamefont
  {Ingeberg}, \citenamefont {Siem}, \citenamefont {Wiedeking}, \citenamefont
  {Sieja}, \citenamefont {Bleuel}, \citenamefont {Brits}, \citenamefont
  {Bucher}, \citenamefont {Dinoko}, \citenamefont {Easton}, \citenamefont
  {rgen}, \citenamefont {Guttormsen}, \citenamefont {Jones}, \citenamefont
  {Kheswa}, \citenamefont {Khumalo}, \citenamefont {Larsen}, \citenamefont
  {Lawrie}, \citenamefont {Lawrie}, \citenamefont {Majola}, \citenamefont
  {Malatji},\ and\ \citenamefont {Makhathini}}]{Ingeberg_2000}%
  \BibitemOpen
  \bibfield  {author} {\bibinfo {author} {\bibfnamefont {V.~W.}\ \bibnamefont
  {Ingeberg}}, \bibinfo {author} {\bibfnamefont {S.}~\bibnamefont {Siem}},
  \bibinfo {author} {\bibfnamefont {M.}~\bibnamefont {Wiedeking}}, \bibinfo
  {author} {\bibfnamefont {K.}~\bibnamefont {Sieja}}, \bibinfo {author}
  {\bibfnamefont {D.~L.}\ \bibnamefont {Bleuel}}, \bibinfo {author}
  {\bibfnamefont {C.}~\bibnamefont {Brits}}, \bibinfo {author} {\bibfnamefont
  {T.~D.}\ \bibnamefont {Bucher}}, \bibinfo {author} {\bibfnamefont {T.~S.}\
  \bibnamefont {Dinoko}}, \bibinfo {author} {\bibfnamefont {J.~L.}\
  \bibnamefont {Easton}}, \bibinfo {author} {\bibfnamefont {A.~G.}\
  \bibnamefont {rgen}}, \bibinfo {author} {\bibfnamefont {M.}~\bibnamefont
  {Guttormsen}}, \bibinfo {author} {\bibfnamefont {P.}~\bibnamefont {Jones}},
  \bibinfo {author} {\bibfnamefont {B.~V.}\ \bibnamefont {Kheswa}}, \bibinfo
  {author} {\bibfnamefont {N.~A.}\ \bibnamefont {Khumalo}}, \bibinfo {author}
  {\bibfnamefont {A.~C.}\ \bibnamefont {Larsen}}, \bibinfo {author}
  {\bibfnamefont {E.~A.}\ \bibnamefont {Lawrie}}, \bibinfo {author}
  {\bibfnamefont {J.~J.}\ \bibnamefont {Lawrie}}, \bibinfo {author}
  {\bibfnamefont {S.~N.~T.}\ \bibnamefont {Majola}}, \bibinfo {author}
  {\bibfnamefont {K.~L.}\ \bibnamefont {Malatji}},\ and\ \bibinfo {author}
  {\bibfnamefont {L.}~\bibnamefont {Makhathini}},\ }\href
  {https://doi.org/https://doi.org/10.1140/epja/s10050-020-00070-7} {\bibfield
  {journal} {\bibinfo  {journal} {The European Physical Journal A}\ }\textbf
  {\bibinfo {volume} {56}},\ \bibinfo {pages} {68} (\bibinfo {year}
  {2000})}\BibitemShut {NoStop}%
\bibitem [{\citenamefont {Ingeberg}\ \emph {et~al.}(2025)\citenamefont
  {Ingeberg}, \citenamefont {Siem}, \citenamefont {Wiedeking}, \citenamefont
  {Choplin}, \citenamefont {Goriely}, \citenamefont {Siess}, \citenamefont
  {Abrahams}, \citenamefont {Arnswald}, \citenamefont {Bello~Garrote},
  \citenamefont {Bleuel}, \citenamefont {Cederk\"all}, \citenamefont
  {Christoffersen}, \citenamefont {Cox}, \citenamefont {De~Witte},
  \citenamefont {Gaffney}, \citenamefont {G\"orgen}, \citenamefont {Henrich},
  \citenamefont {Illana}, \citenamefont {Jones}, \citenamefont {Kheswa},
  \citenamefont {Kr\"oll}, \citenamefont {Majola}, \citenamefont {Malatji},
  \citenamefont {Ojala}, \citenamefont {Pakarinen}, \citenamefont {Rainovski},
  \citenamefont {Reiter}, \citenamefont {von Schmid}, \citenamefont {Seidlitz},
  \citenamefont {Tveten}, \citenamefont {Warr},\ and\ \citenamefont
  {Zeiser}}]{Ingeberg_2025}%
  \BibitemOpen
  \bibfield  {author} {\bibinfo {author} {\bibfnamefont {V.~W.}\ \bibnamefont
  {Ingeberg}}, \bibinfo {author} {\bibfnamefont {S.}~\bibnamefont {Siem}},
  \bibinfo {author} {\bibfnamefont {M.}~\bibnamefont {Wiedeking}}, \bibinfo
  {author} {\bibfnamefont {A.}~\bibnamefont {Choplin}}, \bibinfo {author}
  {\bibfnamefont {S.}~\bibnamefont {Goriely}}, \bibinfo {author} {\bibfnamefont
  {L.}~\bibnamefont {Siess}}, \bibinfo {author} {\bibfnamefont {K.~J.}\
  \bibnamefont {Abrahams}}, \bibinfo {author} {\bibfnamefont {K.}~\bibnamefont
  {Arnswald}}, \bibinfo {author} {\bibfnamefont {F.}~\bibnamefont
  {Bello~Garrote}}, \bibinfo {author} {\bibfnamefont {D.~L.}\ \bibnamefont
  {Bleuel}}, \bibinfo {author} {\bibfnamefont {J.}~\bibnamefont {Cederk\"all}},
  \bibinfo {author} {\bibfnamefont {T.~L.}\ \bibnamefont {Christoffersen}},
  \bibinfo {author} {\bibfnamefont {D.~M.}\ \bibnamefont {Cox}}, \bibinfo
  {author} {\bibfnamefont {H.}~\bibnamefont {De~Witte}}, \bibinfo {author}
  {\bibfnamefont {L.~P.}\ \bibnamefont {Gaffney}}, \bibinfo {author}
  {\bibfnamefont {A.}~\bibnamefont {G\"orgen}}, \bibinfo {author}
  {\bibfnamefont {C.}~\bibnamefont {Henrich}}, \bibinfo {author} {\bibfnamefont
  {A.}~\bibnamefont {Illana}}, \bibinfo {author} {\bibfnamefont
  {P.}~\bibnamefont {Jones}}, \bibinfo {author} {\bibfnamefont {B.~V.}\
  \bibnamefont {Kheswa}}, \bibinfo {author} {\bibfnamefont {T.}~\bibnamefont
  {Kr\"oll}}, \bibinfo {author} {\bibfnamefont {S.~N.~T.}\ \bibnamefont
  {Majola}}, \bibinfo {author} {\bibfnamefont {K.~L.}\ \bibnamefont {Malatji}},
  \bibinfo {author} {\bibfnamefont {J.}~\bibnamefont {Ojala}}, \bibinfo
  {author} {\bibfnamefont {J.}~\bibnamefont {Pakarinen}}, \bibinfo {author}
  {\bibfnamefont {G.}~\bibnamefont {Rainovski}}, \bibinfo {author}
  {\bibfnamefont {P.}~\bibnamefont {Reiter}}, \bibinfo {author} {\bibfnamefont
  {M.}~\bibnamefont {von Schmid}}, \bibinfo {author} {\bibfnamefont
  {M.}~\bibnamefont {Seidlitz}}, \bibinfo {author} {\bibfnamefont {G.~M.}\
  \bibnamefont {Tveten}}, \bibinfo {author} {\bibfnamefont {N.}~\bibnamefont
  {Warr}},\ and\ \bibinfo {author} {\bibfnamefont {F.}~\bibnamefont {Zeiser}}
  (\bibinfo {collaboration} {The ISOLDE Collaboration}),\ }\href
  {https://doi.org/10.1103/PhysRevC.111.015803} {\bibfield  {journal} {\bibinfo
   {journal} {Phys. Rev. C}\ }\textbf {\bibinfo {volume} {111}},\ \bibinfo
  {pages} {015803} (\bibinfo {year} {2025})}\BibitemShut {NoStop}%
\bibitem [{\citenamefont {Spyrou}\ \emph {et~al.}(2014)\citenamefont {Spyrou},
  \citenamefont {Liddick}, \citenamefont {Larsen}, \citenamefont {Guttormsen},
  \citenamefont {Cooper}, \citenamefont {Dombos}, \citenamefont {Morrissey},
  \citenamefont {Naqvi}, \citenamefont {Perdikakis}, \citenamefont {Quinn},
  \citenamefont {Renstr\o~m}, \citenamefont {Rodriguez}, \citenamefont {Simon},
  \citenamefont {Sumithrarachchi},\ and\ \citenamefont {Zegers}}]{Spyrou_2014}%
  \BibitemOpen
  \bibfield  {author} {\bibinfo {author} {\bibfnamefont {A.}~\bibnamefont
  {Spyrou}}, \bibinfo {author} {\bibfnamefont {S.~N.}\ \bibnamefont {Liddick}},
  \bibinfo {author} {\bibfnamefont {A.~C.}\ \bibnamefont {Larsen}}, \bibinfo
  {author} {\bibfnamefont {M.}~\bibnamefont {Guttormsen}}, \bibinfo {author}
  {\bibfnamefont {K.}~\bibnamefont {Cooper}}, \bibinfo {author} {\bibfnamefont
  {A.~C.}\ \bibnamefont {Dombos}}, \bibinfo {author} {\bibfnamefont {D.~J.}\
  \bibnamefont {Morrissey}}, \bibinfo {author} {\bibfnamefont {F.}~\bibnamefont
  {Naqvi}}, \bibinfo {author} {\bibfnamefont {G.}~\bibnamefont {Perdikakis}},
  \bibinfo {author} {\bibfnamefont {S.~J.}\ \bibnamefont {Quinn}}, \bibinfo
  {author} {\bibfnamefont {T.}~\bibnamefont {Renstr\o~m}}, \bibinfo {author}
  {\bibfnamefont {J.~A.}\ \bibnamefont {Rodriguez}}, \bibinfo {author}
  {\bibfnamefont {A.}~\bibnamefont {Simon}}, \bibinfo {author} {\bibfnamefont
  {C.~S.}\ \bibnamefont {Sumithrarachchi}},\ and\ \bibinfo {author}
  {\bibfnamefont {R.~G.~T.}\ \bibnamefont {Zegers}},\ }\href
  {https://doi.org/10.1103/PhysRevLett.113.232502} {\bibfield  {journal}
  {\bibinfo  {journal} {Phys. Rev. Lett.}\ }\textbf {\bibinfo {volume} {113}},\
  \bibinfo {pages} {232502} (\bibinfo {year} {2014})}\BibitemShut {NoStop}%
\bibitem [{\citenamefont {Liddick}\ \emph {et~al.}(2016)\citenamefont
  {Liddick}, \citenamefont {Spyrou}, \citenamefont {Crider}, \citenamefont
  {Naqvi}, \citenamefont {Larsen}, \citenamefont {Guttormsen}, \citenamefont
  {Mumpower}, \citenamefont {Surman}, \citenamefont {Perdikakis}, \citenamefont
  {Bleuel}, \citenamefont {Couture}, \citenamefont {Crespo~Campo},
  \citenamefont {Dombos}, \citenamefont {Lewis}, \citenamefont {Mosby},
  \citenamefont {Nikas}, \citenamefont {Prokop}, \citenamefont {Renstrom},
  \citenamefont {Rubio}, \citenamefont {Siem},\ and\ \citenamefont
  {Quinn}}]{Liddick_2016}%
  \BibitemOpen
  \bibfield  {author} {\bibinfo {author} {\bibfnamefont {S.~N.}\ \bibnamefont
  {Liddick}}, \bibinfo {author} {\bibfnamefont {A.}~\bibnamefont {Spyrou}},
  \bibinfo {author} {\bibfnamefont {B.~P.}\ \bibnamefont {Crider}}, \bibinfo
  {author} {\bibfnamefont {F.}~\bibnamefont {Naqvi}}, \bibinfo {author}
  {\bibfnamefont {A.~C.}\ \bibnamefont {Larsen}}, \bibinfo {author}
  {\bibfnamefont {M.}~\bibnamefont {Guttormsen}}, \bibinfo {author}
  {\bibfnamefont {M.}~\bibnamefont {Mumpower}}, \bibinfo {author}
  {\bibfnamefont {R.}~\bibnamefont {Surman}}, \bibinfo {author} {\bibfnamefont
  {G.}~\bibnamefont {Perdikakis}}, \bibinfo {author} {\bibfnamefont {D.~L.}\
  \bibnamefont {Bleuel}}, \bibinfo {author} {\bibfnamefont {A.}~\bibnamefont
  {Couture}}, \bibinfo {author} {\bibfnamefont {L.}~\bibnamefont
  {Crespo~Campo}}, \bibinfo {author} {\bibfnamefont {A.~C.}\ \bibnamefont
  {Dombos}}, \bibinfo {author} {\bibfnamefont {R.}~\bibnamefont {Lewis}},
  \bibinfo {author} {\bibfnamefont {S.}~\bibnamefont {Mosby}}, \bibinfo
  {author} {\bibfnamefont {S.}~\bibnamefont {Nikas}}, \bibinfo {author}
  {\bibfnamefont {C.~J.}\ \bibnamefont {Prokop}}, \bibinfo {author}
  {\bibfnamefont {T.}~\bibnamefont {Renstrom}}, \bibinfo {author}
  {\bibfnamefont {B.}~\bibnamefont {Rubio}}, \bibinfo {author} {\bibfnamefont
  {S.}~\bibnamefont {Siem}},\ and\ \bibinfo {author} {\bibfnamefont {S.~J.}\
  \bibnamefont {Quinn}},\ }\href
  {https://doi.org/10.1103/PhysRevLett.116.242502} {\bibfield  {journal}
  {\bibinfo  {journal} {Phys. Rev. Lett.}\ }\textbf {\bibinfo {volume} {116}},\
  \bibinfo {pages} {242502} (\bibinfo {year} {2016})}\BibitemShut {NoStop}%
\bibitem [{\citenamefont {Escher}\ \emph {et~al.}(2012)\citenamefont {Escher},
  \citenamefont {Harke}, \citenamefont {Dietrich}, \citenamefont {Scielzo},
  \citenamefont {Thompson},\ and\ \citenamefont {Younes}}]{Escher_2012}%
  \BibitemOpen
  \bibfield  {author} {\bibinfo {author} {\bibfnamefont {J.~E.}\ \bibnamefont
  {Escher}}, \bibinfo {author} {\bibfnamefont {J.~T.}\ \bibnamefont {Harke}},
  \bibinfo {author} {\bibfnamefont {F.~S.}\ \bibnamefont {Dietrich}}, \bibinfo
  {author} {\bibfnamefont {N.~D.}\ \bibnamefont {Scielzo}}, \bibinfo {author}
  {\bibfnamefont {I.~J.}\ \bibnamefont {Thompson}},\ and\ \bibinfo {author}
  {\bibfnamefont {W.}~\bibnamefont {Younes}},\ }\href
  {https://doi.org/10.1103/RevModPhys.84.353} {\bibfield  {journal} {\bibinfo
  {journal} {Rev. Mod. Phys.}\ }\textbf {\bibinfo {volume} {84}},\ \bibinfo
  {pages} {353} (\bibinfo {year} {2012})}\BibitemShut {NoStop}%
\bibitem [{\citenamefont {Escher}\ \emph {et~al.}(2018)\citenamefont {Escher},
  \citenamefont {Harke}, \citenamefont {Hughes}, \citenamefont {Scielzo},
  \citenamefont {Casperson}, \citenamefont {Ota}, \citenamefont {Park},
  \citenamefont {Saastamoinen},\ and\ \citenamefont {Ross}}]{Escher_2018}%
  \BibitemOpen
  \bibfield  {author} {\bibinfo {author} {\bibfnamefont {J.~E.}\ \bibnamefont
  {Escher}}, \bibinfo {author} {\bibfnamefont {J.~T.}\ \bibnamefont {Harke}},
  \bibinfo {author} {\bibfnamefont {R.~O.}\ \bibnamefont {Hughes}}, \bibinfo
  {author} {\bibfnamefont {N.~D.}\ \bibnamefont {Scielzo}}, \bibinfo {author}
  {\bibfnamefont {R.~J.}\ \bibnamefont {Casperson}}, \bibinfo {author}
  {\bibfnamefont {S.}~\bibnamefont {Ota}}, \bibinfo {author} {\bibfnamefont
  {H.~I.}\ \bibnamefont {Park}}, \bibinfo {author} {\bibfnamefont
  {A.}~\bibnamefont {Saastamoinen}},\ and\ \bibinfo {author} {\bibfnamefont
  {T.~J.}\ \bibnamefont {Ross}},\ }\href
  {https://doi.org/10.1103/PhysRevLett.121.052501} {\bibfield  {journal}
  {\bibinfo  {journal} {Phys. Rev. Lett.}\ }\textbf {\bibinfo {volume} {121}},\
  \bibinfo {pages} {052501} (\bibinfo {year} {2018})}\BibitemShut {NoStop}%
\bibitem [{\citenamefont {Ratkiewicz}\ \emph {et~al.}(2019)\citenamefont
  {Ratkiewicz}, \citenamefont {Cizewski}, \citenamefont {Escher}, \citenamefont
  {Potel}, \citenamefont {Harke}, \citenamefont {Casperson}, \citenamefont
  {McCleskey}, \citenamefont {Austin}, \citenamefont {Burcher}, \citenamefont
  {Hughes}, \citenamefont {Manning}, \citenamefont {Pain}, \citenamefont
  {Peters}, \citenamefont {Rice}, \citenamefont {Ross}, \citenamefont
  {Scielzo}, \citenamefont {Shand},\ and\ \citenamefont
  {Smith}}]{Ratkiewicz_2019}%
  \BibitemOpen
  \bibfield  {author} {\bibinfo {author} {\bibfnamefont {A.}~\bibnamefont
  {Ratkiewicz}}, \bibinfo {author} {\bibfnamefont {J.~A.}\ \bibnamefont
  {Cizewski}}, \bibinfo {author} {\bibfnamefont {J.~E.}\ \bibnamefont
  {Escher}}, \bibinfo {author} {\bibfnamefont {G.}~\bibnamefont {Potel}},
  \bibinfo {author} {\bibfnamefont {J.~T.}\ \bibnamefont {Harke}}, \bibinfo
  {author} {\bibfnamefont {R.~J.}\ \bibnamefont {Casperson}}, \bibinfo {author}
  {\bibfnamefont {M.}~\bibnamefont {McCleskey}}, \bibinfo {author}
  {\bibfnamefont {R.~A.~E.}\ \bibnamefont {Austin}}, \bibinfo {author}
  {\bibfnamefont {S.}~\bibnamefont {Burcher}}, \bibinfo {author} {\bibfnamefont
  {R.~O.}\ \bibnamefont {Hughes}}, \bibinfo {author} {\bibfnamefont
  {B.}~\bibnamefont {Manning}}, \bibinfo {author} {\bibfnamefont {S.~D.}\
  \bibnamefont {Pain}}, \bibinfo {author} {\bibfnamefont {W.~A.}\ \bibnamefont
  {Peters}}, \bibinfo {author} {\bibfnamefont {S.}~\bibnamefont {Rice}},
  \bibinfo {author} {\bibfnamefont {T.~J.}\ \bibnamefont {Ross}}, \bibinfo
  {author} {\bibfnamefont {N.~D.}\ \bibnamefont {Scielzo}}, \bibinfo {author}
  {\bibfnamefont {C.}~\bibnamefont {Shand}},\ and\ \bibinfo {author}
  {\bibfnamefont {K.}~\bibnamefont {Smith}},\ }\href
  {https://doi.org/10.1103/PhysRevLett.122.052502} {\bibfield  {journal}
  {\bibinfo  {journal} {Phys. Rev. Lett.}\ }\textbf {\bibinfo {volume} {122}},\
  \bibinfo {pages} {052502} (\bibinfo {year} {2019})}\BibitemShut {NoStop}%
\bibitem [{\citenamefont {Hauser}\ and\ \citenamefont
  {Feshbach}(1952)}]{Hauser}%
  \BibitemOpen
  \bibfield  {author} {\bibinfo {author} {\bibfnamefont {W.}~\bibnamefont
  {Hauser}}\ and\ \bibinfo {author} {\bibfnamefont {H.}~\bibnamefont
  {Feshbach}},\ }\href {https://doi.org/10.1103/PhysRev.87.366} {\bibfield
  {journal} {\bibinfo  {journal} {Phys. Rev.}\ }\textbf {\bibinfo {volume}
  {87}},\ \bibinfo {pages} {366} (\bibinfo {year} {1952})}\BibitemShut
  {NoStop}%
\bibitem [{\citenamefont {Bethe}(1936)}]{Bethe_1936}%
  \BibitemOpen
  \bibfield  {author} {\bibinfo {author} {\bibfnamefont {H.~A.}\ \bibnamefont
  {Bethe}},\ }\href {https://doi.org/10.1103/PhysRev.50.332} {\bibfield
  {journal} {\bibinfo  {journal} {Phys. Rev.}\ }\textbf {\bibinfo {volume}
  {50}},\ \bibinfo {pages} {332} (\bibinfo {year} {1936})}\BibitemShut
  {NoStop}%
\bibitem [{\citenamefont {Capote}\ \emph {et~al.}(2009)\citenamefont {Capote},
  \citenamefont {Herman}, \citenamefont {Obložinský}, \citenamefont {Young},
  \citenamefont {Goriely}, \citenamefont {Belgya}, \citenamefont {Ignatyuk},
  \citenamefont {Koning}, \citenamefont {Hilaire}, \citenamefont {Plujko},
  \citenamefont {Avrigeanu}, \citenamefont {Bersillon}, \citenamefont
  {Chadwick}, \citenamefont {Fukahori}, \citenamefont {Ge}, \citenamefont
  {Han}, \citenamefont {Kailas}, \citenamefont {Kopecky}, \citenamefont
  {Maslov}, \citenamefont {Reffo}, \citenamefont {Sin}, \citenamefont
  {Soukhovitskii},\ and\ \citenamefont {Talou}}]{Capote_2009}%
  \BibitemOpen
  \bibfield  {author} {\bibinfo {author} {\bibfnamefont {R.}~\bibnamefont
  {Capote}}, \bibinfo {author} {\bibfnamefont {M.}~\bibnamefont {Herman}},
  \bibinfo {author} {\bibfnamefont {P.}~\bibnamefont {Obložinský}}, \bibinfo
  {author} {\bibfnamefont {P.}~\bibnamefont {Young}}, \bibinfo {author}
  {\bibfnamefont {S.}~\bibnamefont {Goriely}}, \bibinfo {author} {\bibfnamefont
  {T.}~\bibnamefont {Belgya}}, \bibinfo {author} {\bibfnamefont
  {A.}~\bibnamefont {Ignatyuk}}, \bibinfo {author} {\bibfnamefont
  {A.}~\bibnamefont {Koning}}, \bibinfo {author} {\bibfnamefont
  {S.}~\bibnamefont {Hilaire}}, \bibinfo {author} {\bibfnamefont
  {V.}~\bibnamefont {Plujko}}, \bibinfo {author} {\bibfnamefont
  {M.}~\bibnamefont {Avrigeanu}}, \bibinfo {author} {\bibfnamefont
  {O.}~\bibnamefont {Bersillon}}, \bibinfo {author} {\bibfnamefont
  {M.}~\bibnamefont {Chadwick}}, \bibinfo {author} {\bibfnamefont
  {T.}~\bibnamefont {Fukahori}}, \bibinfo {author} {\bibfnamefont
  {Z.}~\bibnamefont {Ge}}, \bibinfo {author} {\bibfnamefont {Y.}~\bibnamefont
  {Han}}, \bibinfo {author} {\bibfnamefont {S.}~\bibnamefont {Kailas}},
  \bibinfo {author} {\bibfnamefont {J.}~\bibnamefont {Kopecky}}, \bibinfo
  {author} {\bibfnamefont {V.}~\bibnamefont {Maslov}}, \bibinfo {author}
  {\bibfnamefont {G.}~\bibnamefont {Reffo}}, \bibinfo {author} {\bibfnamefont
  {M.}~\bibnamefont {Sin}}, \bibinfo {author} {\bibfnamefont {E.}~\bibnamefont
  {Soukhovitskii}},\ and\ \bibinfo {author} {\bibfnamefont {P.}~\bibnamefont
  {Talou}},\ }\href {https://doi.org/https://doi.org/10.1016/j.nds.2009.10.004}
  {\bibfield  {journal} {\bibinfo  {journal} {Nuclear Data Sheets}\ }\textbf
  {\bibinfo {volume} {110}},\ \bibinfo {pages} {3107} (\bibinfo {year}
  {2009})}\BibitemShut {NoStop}%
\bibitem [{\citenamefont {Gilbert}\ and\ \citenamefont
  {Cameron}(1965)}]{Gilbert_1965}%
  \BibitemOpen
  \bibfield  {author} {\bibinfo {author} {\bibfnamefont {A.}~\bibnamefont
  {Gilbert}}\ and\ \bibinfo {author} {\bibfnamefont {A.~G.~W.}\ \bibnamefont
  {Cameron}},\ }\href {https://doi.org/10.1139/p65-139} {\bibfield  {journal}
  {\bibinfo  {journal} {Canadian Journal of Physics}\ }\textbf {\bibinfo
  {volume} {43}},\ \bibinfo {pages} {1446} (\bibinfo {year}
  {1965})}\BibitemShut {NoStop}%
\bibitem [{\citenamefont {W.~Dilg}\ and\ \citenamefont
  {Uhl}(1973)}]{Dilg_1973}%
  \BibitemOpen
  \bibfield  {author} {\bibinfo {author} {\bibfnamefont {H.~V.}\ \bibnamefont
  {W.~Dilg}, \bibfnamefont {W.~Schantl}}\ and\ \bibinfo {author} {\bibfnamefont
  {M.}~\bibnamefont {Uhl}},\ }\href
  {https://doi.org/https://doi.org/10.1016/0375-9474(73)90196-6} {\bibfield
  {journal} {\bibinfo  {journal} {Nuclear Physics A}\ }\textbf {\bibinfo
  {volume} {217}},\ \bibinfo {pages} {269} (\bibinfo {year}
  {1973})}\BibitemShut {NoStop}%
\bibitem [{\citenamefont {Ignatyuk}\ \emph {et~al.}(1993)\citenamefont
  {Ignatyuk}, \citenamefont {Weil}, \citenamefont {Raman},\ and\ \citenamefont
  {Kahane}}]{Ignatyuk_1993}%
  \BibitemOpen
  \bibfield  {author} {\bibinfo {author} {\bibfnamefont {A.~V.}\ \bibnamefont
  {Ignatyuk}}, \bibinfo {author} {\bibfnamefont {J.~L.}\ \bibnamefont {Weil}},
  \bibinfo {author} {\bibfnamefont {S.}~\bibnamefont {Raman}},\ and\ \bibinfo
  {author} {\bibfnamefont {S.}~\bibnamefont {Kahane}},\ }\href
  {https://doi.org/10.1103/PhysRevC.47.1504} {\bibfield  {journal} {\bibinfo
  {journal} {Phys. Rev. C}\ }\textbf {\bibinfo {volume} {47}},\ \bibinfo
  {pages} {1504} (\bibinfo {year} {1993})}\BibitemShut {NoStop}%
\bibitem [{\citenamefont {Goriely}\ \emph {et~al.}(2001)\citenamefont
  {Goriely}, \citenamefont {Tondeur},\ and\ \citenamefont
  {Pearson}}]{Goriely_2001}%
  \BibitemOpen
  \bibfield  {author} {\bibinfo {author} {\bibfnamefont {S.}~\bibnamefont
  {Goriely}}, \bibinfo {author} {\bibfnamefont {F.}~\bibnamefont {Tondeur}},\
  and\ \bibinfo {author} {\bibfnamefont {J.}~\bibnamefont {Pearson}},\ }\href
  {https://doi.org/https://doi.org/10.1006/adnd.2000.0857} {\bibfield
  {journal} {\bibinfo  {journal} {Atomic Data and Nuclear Data Tables}\
  }\textbf {\bibinfo {volume} {77}},\ \bibinfo {pages} {311} (\bibinfo {year}
  {2001})}\BibitemShut {NoStop}%
\bibitem [{\citenamefont {Goriely}\ \emph {et~al.}(2008)\citenamefont
  {Goriely}, \citenamefont {Hilaire},\ and\ \citenamefont
  {Koning}}]{Goriely_2008}%
  \BibitemOpen
  \bibfield  {author} {\bibinfo {author} {\bibfnamefont {S.}~\bibnamefont
  {Goriely}}, \bibinfo {author} {\bibfnamefont {S.}~\bibnamefont {Hilaire}},\
  and\ \bibinfo {author} {\bibfnamefont {A.~J.}\ \bibnamefont {Koning}},\
  }\href {https://doi.org/10.1103/PhysRevC.78.064307} {\bibfield  {journal}
  {\bibinfo  {journal} {Phys. Rev. C}\ }\textbf {\bibinfo {volume} {78}},\
  \bibinfo {pages} {064307} (\bibinfo {year} {2008})}\BibitemShut {NoStop}%
\bibitem [{\citenamefont {Hilaire}\ \emph {et~al.}(2012)\citenamefont
  {Hilaire}, \citenamefont {Girod}, \citenamefont {Goriely},\ and\
  \citenamefont {Koning}}]{Hilaire_2012}%
  \BibitemOpen
  \bibfield  {author} {\bibinfo {author} {\bibfnamefont {S.}~\bibnamefont
  {Hilaire}}, \bibinfo {author} {\bibfnamefont {M.}~\bibnamefont {Girod}},
  \bibinfo {author} {\bibfnamefont {S.}~\bibnamefont {Goriely}},\ and\ \bibinfo
  {author} {\bibfnamefont {A.~J.}\ \bibnamefont {Koning}},\ }\href
  {https://doi.org/10.1103/PhysRevC.86.064317} {\bibfield  {journal} {\bibinfo
  {journal} {Phys. Rev. C}\ }\textbf {\bibinfo {volume} {86}},\ \bibinfo
  {pages} {064317} (\bibinfo {year} {2012})}\BibitemShut {NoStop}%
\bibitem [{\citenamefont {Koning}\ \emph {et~al.}(2023)\citenamefont {Koning},
  \citenamefont {Hilaire},\ and\ \citenamefont {Goriely}}]{TALYS}%
  \BibitemOpen
  \bibfield  {author} {\bibinfo {author} {\bibfnamefont {A.}~\bibnamefont
  {Koning}}, \bibinfo {author} {\bibfnamefont {S.}~\bibnamefont {Hilaire}},\
  and\ \bibinfo {author} {\bibfnamefont {S.}~\bibnamefont {Goriely}},\ }\href
  {https://doi.org/https://doi.org/10.1140/epja/s10050-023-01034-3} {\bibfield
  {journal} {\bibinfo  {journal} {The European Physical Journal A}\ }\textbf
  {\bibinfo {volume} {59}},\ \bibinfo {pages} {131} (\bibinfo {year}
  {2023})}\BibitemShut {NoStop}%
\bibitem [{\citenamefont {Alhassid}\ \emph {et~al.}(2015)\citenamefont
  {Alhassid}, \citenamefont {Bonett-Matiz}, \citenamefont {Liu},\ and\
  \citenamefont {Nakada}}]{Alhassid_2015}%
  \BibitemOpen
  \bibfield  {author} {\bibinfo {author} {\bibfnamefont {Y.}~\bibnamefont
  {Alhassid}}, \bibinfo {author} {\bibfnamefont {M.}~\bibnamefont
  {Bonett-Matiz}}, \bibinfo {author} {\bibfnamefont {S.}~\bibnamefont {Liu}},\
  and\ \bibinfo {author} {\bibfnamefont {H.}~\bibnamefont {Nakada}},\ }\href
  {https://doi.org/10.1103/PhysRevC.92.024307} {\bibfield  {journal} {\bibinfo
  {journal} {Phys. Rev. C}\ }\textbf {\bibinfo {volume} {92}},\ \bibinfo
  {pages} {024307} (\bibinfo {year} {2015})}\BibitemShut {NoStop}%
\bibitem [{\citenamefont {DeMartini}\ and\ \citenamefont
  {Alhassid}(2025)}]{DeMartini_2025}%
  \BibitemOpen
  \bibfield  {author} {\bibinfo {author} {\bibfnamefont {D.}~\bibnamefont
  {DeMartini}}\ and\ \bibinfo {author} {\bibfnamefont {Y.}~\bibnamefont
  {Alhassid}},\ }\href {https://doi.org/10.1103/PhysRevC.111.034315} {\bibfield
   {journal} {\bibinfo  {journal} {Phys. Rev. C}\ }\textbf {\bibinfo {volume}
  {111}},\ \bibinfo {pages} {034315} (\bibinfo {year} {2025})}\BibitemShut
  {NoStop}%
\bibitem [{\citenamefont {Ormand}\ and\ \citenamefont
  {Brown}(2020)}]{Ormand_2020}%
  \BibitemOpen
  \bibfield  {author} {\bibinfo {author} {\bibfnamefont {W.~E.}\ \bibnamefont
  {Ormand}}\ and\ \bibinfo {author} {\bibfnamefont {B.~A.}\ \bibnamefont
  {Brown}},\ }\href {https://doi.org/10.1103/PhysRevC.102.014315} {\bibfield
  {journal} {\bibinfo  {journal} {Phys. Rev. C}\ }\textbf {\bibinfo {volume}
  {102}},\ \bibinfo {pages} {014315} (\bibinfo {year} {2020})}\BibitemShut
  {NoStop}%
\bibitem [{\citenamefont {Sen'kov}\ and\ \citenamefont
  {Zelevinsky}(2016)}]{Senkov_2016}%
  \BibitemOpen
  \bibfield  {author} {\bibinfo {author} {\bibfnamefont {R.}~\bibnamefont
  {Sen'kov}}\ and\ \bibinfo {author} {\bibfnamefont {V.}~\bibnamefont
  {Zelevinsky}},\ }\href {https://doi.org/10.1103/PhysRevC.93.064304}
  {\bibfield  {journal} {\bibinfo  {journal} {Phys. Rev. C}\ }\textbf {\bibinfo
  {volume} {93}},\ \bibinfo {pages} {064304} (\bibinfo {year}
  {2016})}\BibitemShut {NoStop}%
\bibitem [{\citenamefont {Chang}\ \emph {et~al.}(1971)\citenamefont {Chang},
  \citenamefont {French},\ and\ \citenamefont {Thio}}]{French1}%
  \BibitemOpen
  \bibfield  {author} {\bibinfo {author} {\bibfnamefont {F.}~\bibnamefont
  {Chang}}, \bibinfo {author} {\bibfnamefont {J.}~\bibnamefont {French}},\ and\
  \bibinfo {author} {\bibfnamefont {T.}~\bibnamefont {Thio}},\ }\href
  {https://doi.org/https://doi.org/10.1016/0003-4916(71)90186-2} {\bibfield
  {journal} {\bibinfo  {journal} {Annals of Physics}\ }\textbf {\bibinfo
  {volume} {66}},\ \bibinfo {pages} {137} (\bibinfo {year} {1971})}\BibitemShut
  {NoStop}%
\bibitem [{\citenamefont {French}\ and\ \citenamefont
  {Ratcliff}(1971)}]{French2}%
  \BibitemOpen
  \bibfield  {author} {\bibinfo {author} {\bibfnamefont {J.~B.}\ \bibnamefont
  {French}}\ and\ \bibinfo {author} {\bibfnamefont {K.~F.}\ \bibnamefont
  {Ratcliff}},\ }\href {https://doi.org/10.1103/PhysRevC.3.94} {\bibfield
  {journal} {\bibinfo  {journal} {Phys. Rev. C}\ }\textbf {\bibinfo {volume}
  {3}},\ \bibinfo {pages} {94} (\bibinfo {year} {1971})}\BibitemShut {NoStop}%
\bibitem [{\citenamefont {Ratcliff}(1971)}]{French3}%
  \BibitemOpen
  \bibfield  {author} {\bibinfo {author} {\bibfnamefont {K.~F.}\ \bibnamefont
  {Ratcliff}},\ }\href {https://doi.org/10.1103/PhysRevC.3.117} {\bibfield
  {journal} {\bibinfo  {journal} {Phys. Rev. C}\ }\textbf {\bibinfo {volume}
  {3}},\ \bibinfo {pages} {117} (\bibinfo {year} {1971})}\BibitemShut {NoStop}%
\bibitem [{\citenamefont {Wong}(1986)}]{Wong}%
  \BibitemOpen
  \bibfield  {author} {\bibinfo {author} {\bibfnamefont {S.~S.~M.}\
  \bibnamefont {Wong}},\ }\href@noop {} {\emph {\bibinfo {title} {Nuclear
  Statistical Spectroscopy}}}\ (\bibinfo  {publisher} {Oxford University
  Press},\ \bibinfo {address} {New York},\ \bibinfo {year} {1986})\BibitemShut
  {NoStop}%
\bibitem [{\citenamefont {Horoi}\ \emph
  {et~al.}(2003{\natexlab{a}})\citenamefont {Horoi}, \citenamefont {Kaiser},\
  and\ \citenamefont {Zelevinsky}}]{HoroiM1}%
  \BibitemOpen
  \bibfield  {author} {\bibinfo {author} {\bibfnamefont {M.}~\bibnamefont
  {Horoi}}, \bibinfo {author} {\bibfnamefont {J.}~\bibnamefont {Kaiser}},\ and\
  \bibinfo {author} {\bibfnamefont {V.}~\bibnamefont {Zelevinsky}},\ }\href
  {https://doi.org/10.1103/PhysRevC.67.054309} {\bibfield  {journal} {\bibinfo
  {journal} {Phys. Rev. C}\ }\textbf {\bibinfo {volume} {67}},\ \bibinfo
  {pages} {054309} (\bibinfo {year} {2003}{\natexlab{a}})}\BibitemShut
  {NoStop}%
\bibitem [{\citenamefont {Horoi}\ \emph {et~al.}(2005)\citenamefont {Horoi},
  \citenamefont {Ghita},\ and\ \citenamefont {Zelevinsky}}]{HoroiM2}%
  \BibitemOpen
  \bibfield  {author} {\bibinfo {author} {\bibfnamefont {M.}~\bibnamefont
  {Horoi}}, \bibinfo {author} {\bibfnamefont {M.}~\bibnamefont {Ghita}},\ and\
  \bibinfo {author} {\bibfnamefont {V.}~\bibnamefont {Zelevinsky}},\ }\href
  {https://doi.org/https://doi.org/10.1016/j.nuclphysa.2005.05.029} {\bibfield
  {journal} {\bibinfo  {journal} {Nuclear Physics A}\ }\textbf {\bibinfo
  {volume} {758}},\ \bibinfo {pages} {142} (\bibinfo {year} {2005})},\ \bibinfo
  {note} {nuclei in the Cosmos VIII}\BibitemShut {NoStop}%
\bibitem [{\citenamefont {Horoi}\ \emph {et~al.}(2004)\citenamefont {Horoi},
  \citenamefont {Ghita},\ and\ \citenamefont {Zelevinsky}}]{HoroiM3}%
  \BibitemOpen
  \bibfield  {author} {\bibinfo {author} {\bibfnamefont {M.}~\bibnamefont
  {Horoi}}, \bibinfo {author} {\bibfnamefont {M.}~\bibnamefont {Ghita}},\ and\
  \bibinfo {author} {\bibfnamefont {V.}~\bibnamefont {Zelevinsky}},\ }\href
  {https://doi.org/10.1103/PhysRevC.69.041307} {\bibfield  {journal} {\bibinfo
  {journal} {Phys. Rev. C}\ }\textbf {\bibinfo {volume} {69}},\ \bibinfo
  {pages} {041307} (\bibinfo {year} {2004})}\BibitemShut {NoStop}%
\bibitem [{\citenamefont {Sen'kov}\ and\ \citenamefont {Horoi}(2010)}]{Senkov}%
  \BibitemOpen
  \bibfield  {author} {\bibinfo {author} {\bibfnamefont {R.~A.}\ \bibnamefont
  {Sen'kov}}\ and\ \bibinfo {author} {\bibfnamefont {M.}~\bibnamefont
  {Horoi}},\ }\href {https://doi.org/10.1103/PhysRevC.82.024304} {\bibfield
  {journal} {\bibinfo  {journal} {Phys. Rev. C}\ }\textbf {\bibinfo {volume}
  {82}},\ \bibinfo {pages} {024304} (\bibinfo {year} {2010})}\BibitemShut
  {NoStop}%
\bibitem [{\citenamefont {Scott}\ and\ \citenamefont {Horoi}(2010)}]{Scott}%
  \BibitemOpen
  \bibfield  {author} {\bibinfo {author} {\bibfnamefont {M.}~\bibnamefont
  {Scott}}\ and\ \bibinfo {author} {\bibfnamefont {M.}~\bibnamefont {Horoi}},\
  }\href {https://doi.org/10.1209/0295-5075/91/52001} {\bibfield  {journal}
  {\bibinfo  {journal} {Europhysics Letters}\ }\textbf {\bibinfo {volume}
  {91}},\ \bibinfo {pages} {52001} (\bibinfo {year} {2010})}\BibitemShut
  {NoStop}%
\bibitem [{\citenamefont {Sen’kov}\ \emph {et~al.}(2013)\citenamefont
  {Sen’kov}, \citenamefont {Horoi},\ and\ \citenamefont
  {Zelevinsky}}]{Senkov1}%
  \BibitemOpen
  \bibfield  {author} {\bibinfo {author} {\bibfnamefont {R.}~\bibnamefont
  {Sen’kov}}, \bibinfo {author} {\bibfnamefont {M.}~\bibnamefont {Horoi}},\
  and\ \bibinfo {author} {\bibfnamefont {V.}~\bibnamefont {Zelevinsky}},\
  }\href {https://doi.org/https://doi.org/10.1016/j.cpc.2012.09.006} {\bibfield
   {journal} {\bibinfo  {journal} {Computer Physics Communications}\ }\textbf
  {\bibinfo {volume} {184}},\ \bibinfo {pages} {215} (\bibinfo {year}
  {2013})}\BibitemShut {NoStop}%
\bibitem [{\citenamefont {Brody}\ \emph {et~al.}(1981)\citenamefont {Brody},
  \citenamefont {Flores}, \citenamefont {French}, \citenamefont {Mello},
  \citenamefont {Pandey},\ and\ \citenamefont {Wong}}]{Brody}%
  \BibitemOpen
  \bibfield  {author} {\bibinfo {author} {\bibfnamefont {T.~A.}\ \bibnamefont
  {Brody}}, \bibinfo {author} {\bibfnamefont {J.}~\bibnamefont {Flores}},
  \bibinfo {author} {\bibfnamefont {J.~B.}\ \bibnamefont {French}}, \bibinfo
  {author} {\bibfnamefont {P.~A.}\ \bibnamefont {Mello}}, \bibinfo {author}
  {\bibfnamefont {A.}~\bibnamefont {Pandey}},\ and\ \bibinfo {author}
  {\bibfnamefont {S.~S.~M.}\ \bibnamefont {Wong}},\ }\href
  {https://doi.org/10.1103/RevModPhys.53.385} {\bibfield  {journal} {\bibinfo
  {journal} {Rev. Mod. Phys.}\ }\textbf {\bibinfo {volume} {53}},\ \bibinfo
  {pages} {385} (\bibinfo {year} {1981})}\BibitemShut {NoStop}%
\bibitem [{\citenamefont {Horoi}\ \emph {et~al.}(1999)\citenamefont {Horoi},
  \citenamefont {Volya},\ and\ \citenamefont {Zelevinsky}}]{Horoi1}%
  \BibitemOpen
  \bibfield  {author} {\bibinfo {author} {\bibfnamefont {M.}~\bibnamefont
  {Horoi}}, \bibinfo {author} {\bibfnamefont {A.}~\bibnamefont {Volya}},\ and\
  \bibinfo {author} {\bibfnamefont {V.}~\bibnamefont {Zelevinsky}},\ }\href
  {https://doi.org/10.1103/PhysRevLett.82.2064} {\bibfield  {journal} {\bibinfo
   {journal} {Phys. Rev. Lett.}\ }\textbf {\bibinfo {volume} {82}},\ \bibinfo
  {pages} {2064} (\bibinfo {year} {1999})}\BibitemShut {NoStop}%
\bibitem [{\citenamefont {Horoi}\ \emph {et~al.}(2002)\citenamefont {Horoi},
  \citenamefont {Alex~Brown},\ and\ \citenamefont {Zelevinsky}}]{Horoi2}%
  \BibitemOpen
  \bibfield  {author} {\bibinfo {author} {\bibfnamefont {M.}~\bibnamefont
  {Horoi}}, \bibinfo {author} {\bibfnamefont {B.}~\bibnamefont {Alex~Brown}},\
  and\ \bibinfo {author} {\bibfnamefont {V.}~\bibnamefont {Zelevinsky}},\
  }\href {https://doi.org/10.1103/PhysRevC.65.027303} {\bibfield  {journal}
  {\bibinfo  {journal} {Phys. Rev. C}\ }\textbf {\bibinfo {volume} {65}},\
  \bibinfo {pages} {027303} (\bibinfo {year} {2002})}\BibitemShut {NoStop}%
\bibitem [{\citenamefont {Horoi}\ \emph
  {et~al.}(2003{\natexlab{b}})\citenamefont {Horoi}, \citenamefont {Brown},\
  and\ \citenamefont {Zelevinsky}}]{Horoi3}%
  \BibitemOpen
  \bibfield  {author} {\bibinfo {author} {\bibfnamefont {M.}~\bibnamefont
  {Horoi}}, \bibinfo {author} {\bibfnamefont {B.~A.}\ \bibnamefont {Brown}},\
  and\ \bibinfo {author} {\bibfnamefont {V.}~\bibnamefont {Zelevinsky}},\
  }\href {https://doi.org/10.1103/PhysRevC.67.034303} {\bibfield  {journal}
  {\bibinfo  {journal} {Phys. Rev. C}\ }\textbf {\bibinfo {volume} {67}},\
  \bibinfo {pages} {034303} (\bibinfo {year} {2003}{\natexlab{b}})}\BibitemShut
  {NoStop}%
\bibitem [{\citenamefont {Horoi}\ and\ \citenamefont
  {Zelevinsky}(2007)}]{Horoi4}%
  \BibitemOpen
  \bibfield  {author} {\bibinfo {author} {\bibfnamefont {M.}~\bibnamefont
  {Horoi}}\ and\ \bibinfo {author} {\bibfnamefont {V.}~\bibnamefont
  {Zelevinsky}},\ }\href {https://doi.org/10.1103/PhysRevLett.98.262503}
  {\bibfield  {journal} {\bibinfo  {journal} {Phys. Rev. Lett.}\ }\textbf
  {\bibinfo {volume} {98}},\ \bibinfo {pages} {262503} (\bibinfo {year}
  {2007})}\BibitemShut {NoStop}%
\bibitem [{\citenamefont {Honma}\ \emph {et~al.}(2004)\citenamefont {Honma},
  \citenamefont {Otsuka}, \citenamefont {Brown},\ and\ \citenamefont
  {Mizusaki}}]{Honma2004}%
  \BibitemOpen
  \bibfield  {author} {\bibinfo {author} {\bibfnamefont {M.}~\bibnamefont
  {Honma}}, \bibinfo {author} {\bibfnamefont {T.}~\bibnamefont {Otsuka}},
  \bibinfo {author} {\bibfnamefont {B.~A.}\ \bibnamefont {Brown}},\ and\
  \bibinfo {author} {\bibfnamefont {T.}~\bibnamefont {Mizusaki}},\ }\href
  {https://doi.org/10.1103/PhysRevC.69.034335} {\bibfield  {journal} {\bibinfo
  {journal} {Phys. Rev. C}\ }\textbf {\bibinfo {volume} {69}},\ \bibinfo
  {pages} {034335} (\bibinfo {year} {2004})}\BibitemShut {NoStop}%
\bibitem [{\citenamefont {Honma}\ \emph {et~al.}(2005)\citenamefont {Honma},
  \citenamefont {Otsuka}, \citenamefont {Brown},\ and\ \citenamefont
  {Mizusaki}}]{Honma2005}%
  \BibitemOpen
  \bibfield  {author} {\bibinfo {author} {\bibfnamefont {M.}~\bibnamefont
  {Honma}}, \bibinfo {author} {\bibfnamefont {T.}~\bibnamefont {Otsuka}},
  \bibinfo {author} {\bibfnamefont {B.~A.}\ \bibnamefont {Brown}},\ and\
  \bibinfo {author} {\bibfnamefont {T.}~\bibnamefont {Mizusaki}},\ }\href
  {https://doi.org/10.1140/epjad/i2005-06-032-2} {\bibfield  {journal}
  {\bibinfo  {journal} {Eur. Phys. J. A}\ }\textbf {\bibinfo {volume} {25}},\
  \bibinfo {pages} {499} (\bibinfo {year} {2005})}\BibitemShut {NoStop}%
\bibitem [{\citenamefont {Brown}\ and\ \citenamefont {Rae}(2014)}]{NuShellX}%
  \BibitemOpen
  \bibfield  {author} {\bibinfo {author} {\bibfnamefont {B.}~\bibnamefont
  {Brown}}\ and\ \bibinfo {author} {\bibfnamefont {W.}~\bibnamefont {Rae}},\
  }\href {https://doi.org/https://doi.org/10.1016/j.nds.2014.07.022} {\bibfield
   {journal} {\bibinfo  {journal} {Nuclear Data Sheets}\ }\textbf {\bibinfo
  {volume} {120}},\ \bibinfo {pages} {115} (\bibinfo {year}
  {2014})}\BibitemShut {NoStop}%
\bibitem [{\citenamefont {Wang}\ \emph {et~al.}(2012)\citenamefont {Wang},
  \citenamefont {Audi}, \citenamefont {Wapstra}, \citenamefont {Kondev},
  \citenamefont {MacCormick}, \citenamefont {Xu},\ and\ \citenamefont
  {Pfeiffer}}]{NNDCFe}%
  \BibitemOpen
  \bibfield  {author} {\bibinfo {author} {\bibfnamefont {M.}~\bibnamefont
  {Wang}}, \bibinfo {author} {\bibfnamefont {G.}~\bibnamefont {Audi}}, \bibinfo
  {author} {\bibfnamefont {A.~H.}\ \bibnamefont {Wapstra}}, \bibinfo {author}
  {\bibfnamefont {F.~G.}\ \bibnamefont {Kondev}}, \bibinfo {author}
  {\bibfnamefont {M.}~\bibnamefont {MacCormick}}, \bibinfo {author}
  {\bibfnamefont {X.}~\bibnamefont {Xu}},\ and\ \bibinfo {author}
  {\bibfnamefont {B.}~\bibnamefont {Pfeiffer}},\ }\href
  {https://doi.org/10.1088/1674-1137/36/12/003} {\bibfield  {journal} {\bibinfo
   {journal} {Chinese Physics C}\ }\textbf {\bibinfo {volume} {36}},\ \bibinfo
  {pages} {1603} (\bibinfo {year} {2012})}\BibitemShut {NoStop}%
\bibitem [{\citenamefont {Algin}\ \emph {et~al.}(2008)\citenamefont {Algin},
  \citenamefont {Agvaanluvsan}, \citenamefont {Guttormsen}, \citenamefont
  {Larsen}, \citenamefont {Mitchell}, \citenamefont {Rekstad}, \citenamefont
  {Schiller}, \citenamefont {Siem},\ and\ \citenamefont {Voinov}}]{Algin}%
  \BibitemOpen
  \bibfield  {author} {\bibinfo {author} {\bibfnamefont {E.}~\bibnamefont
  {Algin}}, \bibinfo {author} {\bibfnamefont {U.}~\bibnamefont {Agvaanluvsan}},
  \bibinfo {author} {\bibfnamefont {M.}~\bibnamefont {Guttormsen}}, \bibinfo
  {author} {\bibfnamefont {A.~C.}\ \bibnamefont {Larsen}}, \bibinfo {author}
  {\bibfnamefont {G.~E.}\ \bibnamefont {Mitchell}}, \bibinfo {author}
  {\bibfnamefont {J.}~\bibnamefont {Rekstad}}, \bibinfo {author} {\bibfnamefont
  {A.}~\bibnamefont {Schiller}}, \bibinfo {author} {\bibfnamefont
  {S.}~\bibnamefont {Siem}},\ and\ \bibinfo {author} {\bibfnamefont
  {A.}~\bibnamefont {Voinov}},\ }\href
  {https://doi.org/10.1103/PhysRevC.78.054321} {\bibfield  {journal} {\bibinfo
  {journal} {Phys. Rev. C}\ }\textbf {\bibinfo {volume} {78}},\ \bibinfo
  {pages} {054321} (\bibinfo {year} {2008})}\BibitemShut {NoStop}%
\bibitem [{\citenamefont {Larsen}\ \emph {et~al.}(2017)\citenamefont {Larsen},
  \citenamefont {Guttormsen}, \citenamefont {Blasi}, \citenamefont {Bracco},
  \citenamefont {Camera}, \citenamefont {Campo}, \citenamefont {Eriksen},
  \citenamefont {Görgen}, \citenamefont {Hagen}, \citenamefont {Ingeberg},
  \citenamefont {Kheswa}, \citenamefont {Leoni}, \citenamefont {Midtbø},
  \citenamefont {Million}, \citenamefont {Nyhus}, \citenamefont {Renstrøm},
  \citenamefont {Rose}, \citenamefont {Ruud}, \citenamefont {Siem},
  \citenamefont {Tornyi}, \citenamefont {Tveten}, \citenamefont {Voinov},
  \citenamefont {Wiedeking},\ and\ \citenamefont {Zeiser}}]{Larsen}%
  \BibitemOpen
  \bibfield  {author} {\bibinfo {author} {\bibfnamefont {A.~C.}\ \bibnamefont
  {Larsen}}, \bibinfo {author} {\bibfnamefont {M.}~\bibnamefont {Guttormsen}},
  \bibinfo {author} {\bibfnamefont {N.}~\bibnamefont {Blasi}}, \bibinfo
  {author} {\bibfnamefont {A.}~\bibnamefont {Bracco}}, \bibinfo {author}
  {\bibfnamefont {F.}~\bibnamefont {Camera}}, \bibinfo {author} {\bibfnamefont
  {L.~C.}\ \bibnamefont {Campo}}, \bibinfo {author} {\bibfnamefont {T.~K.}\
  \bibnamefont {Eriksen}}, \bibinfo {author} {\bibfnamefont {A.}~\bibnamefont
  {Görgen}}, \bibinfo {author} {\bibfnamefont {T.~W.}\ \bibnamefont {Hagen}},
  \bibinfo {author} {\bibfnamefont {V.~W.}\ \bibnamefont {Ingeberg}}, \bibinfo
  {author} {\bibfnamefont {B.~V.}\ \bibnamefont {Kheswa}}, \bibinfo {author}
  {\bibfnamefont {S.}~\bibnamefont {Leoni}}, \bibinfo {author} {\bibfnamefont
  {J.~E.}\ \bibnamefont {Midtbø}}, \bibinfo {author} {\bibfnamefont
  {B.}~\bibnamefont {Million}}, \bibinfo {author} {\bibfnamefont {H.~T.}\
  \bibnamefont {Nyhus}}, \bibinfo {author} {\bibfnamefont {T.}~\bibnamefont
  {Renstrøm}}, \bibinfo {author} {\bibfnamefont {S.~J.}\ \bibnamefont {Rose}},
  \bibinfo {author} {\bibfnamefont {I.~E.}\ \bibnamefont {Ruud}}, \bibinfo
  {author} {\bibfnamefont {S.}~\bibnamefont {Siem}}, \bibinfo {author}
  {\bibfnamefont {T.~G.}\ \bibnamefont {Tornyi}}, \bibinfo {author}
  {\bibfnamefont {G.~M.}\ \bibnamefont {Tveten}}, \bibinfo {author}
  {\bibfnamefont {A.~V.}\ \bibnamefont {Voinov}}, \bibinfo {author}
  {\bibfnamefont {M.}~\bibnamefont {Wiedeking}},\ and\ \bibinfo {author}
  {\bibfnamefont {F.}~\bibnamefont {Zeiser}},\ }\href
  {https://doi.org/10.1088/1361-6471/aa644a} {\bibfield  {journal} {\bibinfo
  {journal} {Journal of Physics G: Nuclear and Particle Physics}\ }\textbf
  {\bibinfo {volume} {44}},\ \bibinfo {pages} {064005} (\bibinfo {year}
  {2017})}\BibitemShut {NoStop}%
\bibitem [{\citenamefont {Larsen}(2017)}]{Larsen_priv}%
  \BibitemOpen
  \bibfield  {author} {\bibinfo {author} {\bibfnamefont {A.~C.}\ \bibnamefont
  {Larsen}},\ }\href@noop {} {}\bibinfo {howpublished} {private communication}
  (\bibinfo {year} {2017})\BibitemShut {NoStop}%
\bibitem [{\citenamefont {Mughabghab}(2018)}]{Mughabghab}%
  \BibitemOpen
  \bibfield  {author} {\bibinfo {author} {\bibfnamefont {S.}~\bibnamefont
  {Mughabghab}},\ }\href@noop {} {\emph {\bibinfo {title} {Atlas of Neutron
  Resonances, Volume 1: Resonance Properties and Thermal Cross Sections Z =
  1–60}}},\ \bibinfo {edition} {6th}\ ed.\ (\bibinfo  {publisher}
  {Elsevier},\ \bibinfo {address} {Amsterdam},\ \bibinfo {year}
  {2018})\BibitemShut {NoStop}%
\bibitem [{\citenamefont {Sangeeta}\ \emph {et~al.}(2022)\citenamefont
  {Sangeeta}, \citenamefont {Ghosh}, \citenamefont {Maheshwari}, \citenamefont
  {Saxena},\ and\ \citenamefont {Agrawal}}]{Sangeeta_2022}%
  \BibitemOpen
  \bibfield  {author} {\bibinfo {author} {\bibnamefont {Sangeeta}}, \bibinfo
  {author} {\bibfnamefont {T.}~\bibnamefont {Ghosh}}, \bibinfo {author}
  {\bibfnamefont {B.}~\bibnamefont {Maheshwari}}, \bibinfo {author}
  {\bibfnamefont {G.}~\bibnamefont {Saxena}},\ and\ \bibinfo {author}
  {\bibfnamefont {B.~K.}\ \bibnamefont {Agrawal}},\ }\bibfield  {title}
  {\bibinfo {title} {Astrophysical reaction rates with realistic nuclear level
  densities},\ }\href {https://doi.org/10.1103/PhysRevC.105.044320} {\bibfield
  {journal} {\bibinfo  {journal} {Phys. Rev. C}\ }\textbf {\bibinfo {volume}
  {105}},\ \bibinfo {pages} {044320} (\bibinfo {year} {2022})}\BibitemShut
  {NoStop}%
\bibitem [{\citenamefont {Brink}(1957)}]{Brink_1957}%
  \BibitemOpen
  \bibfield  {author} {\bibinfo {author} {\bibfnamefont {D.}~\bibnamefont
  {Brink}},\ }\href
  {https://doi.org/https://doi.org/10.1016/0029-5582(87)90021-6} {\bibfield
  {journal} {\bibinfo  {journal} {Nuclear Physics}\ }\textbf {\bibinfo {volume}
  {4}},\ \bibinfo {pages} {215} (\bibinfo {year} {1957})}\BibitemShut {NoStop}%
\bibitem [{\citenamefont {Axel}(1962)}]{Axel_1962}%
  \BibitemOpen
  \bibfield  {author} {\bibinfo {author} {\bibfnamefont {P.}~\bibnamefont
  {Axel}},\ }\href {https://doi.org/10.1103/PhysRev.126.671} {\bibfield
  {journal} {\bibinfo  {journal} {Phys. Rev.}\ }\textbf {\bibinfo {volume}
  {126}},\ \bibinfo {pages} {671} (\bibinfo {year} {1962})}\BibitemShut
  {NoStop}%
\bibitem [{\citenamefont {Kopecky}\ and\ \citenamefont
  {Uhl}(1990)}]{Kopecky_1941}%
  \BibitemOpen
  \bibfield  {author} {\bibinfo {author} {\bibfnamefont {J.}~\bibnamefont
  {Kopecky}}\ and\ \bibinfo {author} {\bibfnamefont {M.}~\bibnamefont {Uhl}},\
  }\bibfield  {title} {\bibinfo {title} {Test of gamma-ray strength functions
  in nuclear reaction model calculations},\ }\href
  {https://doi.org/10.1103/PhysRevC.41.1941} {\bibfield  {journal} {\bibinfo
  {journal} {Phys. Rev. C}\ }\textbf {\bibinfo {volume} {41}},\ \bibinfo
  {pages} {1941} (\bibinfo {year} {1990})}\BibitemShut {NoStop}%
\bibitem [{\citenamefont {Goriely}(1998)}]{Goriely_1998}%
  \BibitemOpen
  \bibfield  {author} {\bibinfo {author} {\bibfnamefont {S.}~\bibnamefont
  {Goriely}},\ }\href
  {https://doi.org/https://doi.org/10.1016/S0370-2693(98)00907-1} {\bibfield
  {journal} {\bibinfo  {journal} {Physics Letters B}\ }\textbf {\bibinfo
  {volume} {436}},\ \bibinfo {pages} {10} (\bibinfo {year} {1998})}\BibitemShut
  {NoStop}%
\bibitem [{\citenamefont {Goriely}\ and\ \citenamefont
  {Khan}(2002)}]{Goriely_2002}%
  \BibitemOpen
  \bibfield  {author} {\bibinfo {author} {\bibfnamefont {S.}~\bibnamefont
  {Goriely}}\ and\ \bibinfo {author} {\bibfnamefont {E.}~\bibnamefont {Khan}},\
  }\href {https://doi.org/https://doi.org/10.1016/S0375-9474(02)00860-6}
  {\bibfield  {journal} {\bibinfo  {journal} {Nuclear Physics A}\ }\textbf
  {\bibinfo {volume} {706}},\ \bibinfo {pages} {217} (\bibinfo {year}
  {2002})}\BibitemShut {NoStop}%
\bibitem [{\citenamefont {Goriely}\ \emph {et~al.}(2004)\citenamefont
  {Goriely}, \citenamefont {Khan},\ and\ \citenamefont {Samyn}}]{Goriely_2004}%
  \BibitemOpen
  \bibfield  {author} {\bibinfo {author} {\bibfnamefont {S.}~\bibnamefont
  {Goriely}}, \bibinfo {author} {\bibfnamefont {E.}~\bibnamefont {Khan}},\ and\
  \bibinfo {author} {\bibfnamefont {M.}~\bibnamefont {Samyn}},\ }\href
  {https://doi.org/https://doi.org/10.1016/j.nuclphysa.2004.04.105} {\bibfield
  {journal} {\bibinfo  {journal} {Nuclear Physics A}\ }\textbf {\bibinfo
  {volume} {739}},\ \bibinfo {pages} {331} (\bibinfo {year}
  {2004})}\BibitemShut {NoStop}%
\bibitem [{\citenamefont {Daoutidis}\ and\ \citenamefont
  {Goriely}(2012)}]{Daoutidis_2012}%
  \BibitemOpen
  \bibfield  {author} {\bibinfo {author} {\bibfnamefont {I.}~\bibnamefont
  {Daoutidis}}\ and\ \bibinfo {author} {\bibfnamefont {S.}~\bibnamefont
  {Goriely}},\ }\bibfield  {title} {\bibinfo {title} {Large-scale continuum
  random-phase approximation predictions of dipole strength for astrophysical
  applications},\ }\href {https://doi.org/10.1103/PhysRevC.86.034328}
  {\bibfield  {journal} {\bibinfo  {journal} {Phys. Rev. C}\ }\textbf {\bibinfo
  {volume} {86}},\ \bibinfo {pages} {034328} (\bibinfo {year}
  {2012})}\BibitemShut {NoStop}%
\bibitem [{\citenamefont {Goriely}\ \emph {et~al.}(2018)\citenamefont
  {Goriely}, \citenamefont {Hilaire}, \citenamefont {P\'eru},\ and\
  \citenamefont {Sieja}}]{Goriely_2018}%
  \BibitemOpen
  \bibfield  {author} {\bibinfo {author} {\bibfnamefont {S.}~\bibnamefont
  {Goriely}}, \bibinfo {author} {\bibfnamefont {S.}~\bibnamefont {Hilaire}},
  \bibinfo {author} {\bibfnamefont {S.}~\bibnamefont {P\'eru}},\ and\ \bibinfo
  {author} {\bibfnamefont {K.}~\bibnamefont {Sieja}},\ }\bibfield  {title}
  {\bibinfo {title} {Gogny-hfb+qrpa dipole strength function and its
  application to radiative nucleon capture cross section},\ }\href
  {https://doi.org/10.1103/PhysRevC.98.014327} {\bibfield  {journal} {\bibinfo
  {journal} {Phys. Rev. C}\ }\textbf {\bibinfo {volume} {98}},\ \bibinfo
  {pages} {014327} (\bibinfo {year} {2018})}\BibitemShut {NoStop}%
\bibitem [{\citenamefont {Koning}\ and\ \citenamefont
  {Delaroche}(2003)}]{Koning_2003}%
  \BibitemOpen
  \bibfield  {author} {\bibinfo {author} {\bibfnamefont {A.}~\bibnamefont
  {Koning}}\ and\ \bibinfo {author} {\bibfnamefont {J.}~\bibnamefont
  {Delaroche}},\ }\href
  {https://doi.org/https://doi.org/10.1016/S0375-9474(02)01321-0} {\bibfield
  {journal} {\bibinfo  {journal} {Nuclear Physics A}\ }\textbf {\bibinfo
  {volume} {713}},\ \bibinfo {pages} {231} (\bibinfo {year}
  {2003})}\BibitemShut {NoStop}%
\bibitem [{\citenamefont {Brown}\ \emph {et~al.}(2018)\citenamefont {Brown},
  \citenamefont {Chadwick}, \citenamefont {Capote} \emph {et~al.}}]{Brown2018}%
  \BibitemOpen
  \bibfield  {author} {\bibinfo {author} {\bibfnamefont {D.}~\bibnamefont
  {Brown}}, \bibinfo {author} {\bibfnamefont {M.}~\bibnamefont {Chadwick}},
  \bibinfo {author} {\bibfnamefont {R.}~\bibnamefont {Capote}}, \emph
  {et~al.},\ }\bibfield  {title} {\bibinfo {title} {{ENDF/B-VIII.0}: The
  {8$^{th}$} major release of the nuclear reaction data library with
  {CIELO}-project cross sections, new standards and thermal scattering data},\
  }\href {https://doi.org/https://doi.org/10.1016/j.nds.2018.02.001} {\bibfield
   {journal} {\bibinfo  {journal} {Nuclear Data Sheets}\ }\textbf {\bibinfo
  {volume} {148}},\ \bibinfo {pages} {1 } (\bibinfo {year} {2018})},\ \bibinfo
  {note} {special Issue on Nuclear Reaction Data}\BibitemShut {NoStop}%
\bibitem [{\citenamefont {Dillmann}\ \emph {et~al.}(2014)\citenamefont
  {Dillmann}, \citenamefont {Szücs}, \citenamefont {Plag}, \citenamefont
  {Fülöp}, \citenamefont {Käppeler}, \citenamefont {Mengoni},\ and\
  \citenamefont {Rauscher}}]{Kadonis}%
  \BibitemOpen
  \bibfield  {author} {\bibinfo {author} {\bibfnamefont {I.}~\bibnamefont
  {Dillmann}}, \bibinfo {author} {\bibfnamefont {T.}~\bibnamefont {Szücs}},
  \bibinfo {author} {\bibfnamefont {R.}~\bibnamefont {Plag}}, \bibinfo {author}
  {\bibfnamefont {Z.}~\bibnamefont {Fülöp}}, \bibinfo {author} {\bibfnamefont
  {F.}~\bibnamefont {Käppeler}}, \bibinfo {author} {\bibfnamefont
  {A.}~\bibnamefont {Mengoni}},\ and\ \bibinfo {author} {\bibfnamefont
  {T.}~\bibnamefont {Rauscher}},\ }\href
  {https://doi.org/https://doi.org/10.1016/j.nds.2014.07.038} {\bibfield
  {journal} {\bibinfo  {journal} {Nuclear Data Sheets}\ }\textbf {\bibinfo
  {volume} {120}},\ \bibinfo {pages} {171} (\bibinfo {year}
  {2014})}\BibitemShut {NoStop}%
\bibitem [{\citenamefont {Cyburt}\ \emph {et~al.}(2010)\citenamefont {Cyburt},
  \citenamefont {Amthor}, \citenamefont {Ferguson}, \citenamefont {Meisel},
  \citenamefont {Smith}, \citenamefont {Warren}, \citenamefont {Heger},
  \citenamefont {Hoffman}, \citenamefont {Rauscher}, \citenamefont {Sakharuk},
  \citenamefont {Schatz}, \citenamefont {Thielemann},\ and\ \citenamefont
  {Wiescher}}]{JINA}%
  \BibitemOpen
  \bibfield  {author} {\bibinfo {author} {\bibfnamefont {R.~H.}\ \bibnamefont
  {Cyburt}}, \bibinfo {author} {\bibfnamefont {A.~M.}\ \bibnamefont {Amthor}},
  \bibinfo {author} {\bibfnamefont {R.}~\bibnamefont {Ferguson}}, \bibinfo
  {author} {\bibfnamefont {Z.}~\bibnamefont {Meisel}}, \bibinfo {author}
  {\bibfnamefont {K.}~\bibnamefont {Smith}}, \bibinfo {author} {\bibfnamefont
  {S.}~\bibnamefont {Warren}}, \bibinfo {author} {\bibfnamefont
  {A.}~\bibnamefont {Heger}}, \bibinfo {author} {\bibfnamefont {R.~D.}\
  \bibnamefont {Hoffman}}, \bibinfo {author} {\bibfnamefont {T.}~\bibnamefont
  {Rauscher}}, \bibinfo {author} {\bibfnamefont {A.}~\bibnamefont {Sakharuk}},
  \bibinfo {author} {\bibfnamefont {H.}~\bibnamefont {Schatz}}, \bibinfo
  {author} {\bibfnamefont {F.~K.}\ \bibnamefont {Thielemann}},\ and\ \bibinfo
  {author} {\bibfnamefont {M.}~\bibnamefont {Wiescher}},\ }\href
  {https://doi.org/10.1088/0067-0049/189/1/240} {\bibfield  {journal} {\bibinfo
   {journal} {The Astrophysical Journal Supplement Series}\ }\textbf {\bibinfo
  {volume} {189}},\ \bibinfo {pages} {240} (\bibinfo {year}
  {2010})}\BibitemShut {NoStop}%
\end{thebibliography}%

\end{document}